\documentclass[sigconf]{acmart}
\usepackage{makecell}
\usepackage{multirow}
\usepackage{subfigure}
\AtBeginDocument{%
	\providecommand\BibTeX{{%
			\normalfont B\kern-0.5em{\scshape i\kern-0.25em b}\kern-0.8em\TeX}}}

\copyrightyear{2022}
\acmYear{2022}
\setcopyright{rightsretained}
\acmConference[WWW '22]{Proceedings of the ACM Web Conference 2022}{April 25--29, 2022}{Lyon, France}
\acmBooktitle{Proceedings of the ACM Web Conference 2022 (WWW '22), April 25--29, 2022, Virtual Event, Lyon, France}\acmDOI{10.1145/3485447.3512217}
\acmISBN{978-1-4503-9096-5/22/04}

\begin{document}
	\title{ET-BERT: A Contextualized Datagram Representation with Pre-training Transformers for Encrypted Traffic Classification}
	
	
	\author{Xinjie Lin$^{1,2}$, Gang Xiong$^{1,2}$, Gaopeng Gou$^{1,2}$, Zhen Li$^{1,2}$, Junzheng Shi$^1$, Jing Yu$^{1,2}$$^{\ast}$}
	\thanks{$\ast$ Corresponding Author}
	\affiliation{%
		$^1$Institute of Information Engineering, Chinese Academy of Sciences
		\city{Beijing}
		\country{China}
	}
	\affiliation{$^2$School of Cyber Security, University of the Chinese Academy of Sciences
		\city{Beijing}
		\country{China}
	}
	
	
	\renewcommand{\shortauthors}{Xinjie Lin et al.}
	
	\begin{abstract}
		Encrypted traffic classification requires discriminative and robust traffic representation captured from content-invisible and imbalanced traffic data for accurate classification, which is challenging but indispensable to achieve network security and network management. The major limitation of existing solutions is that they highly rely on the deep features, which are overly dependent on data size and hard to generalize on unseen data. How to leverage the open-domain unlabeled traffic data to learn representation with strong generalization ability remains a key challenge. In this paper, we propose a new traffic representation model called \textbf{E}ncrypted \textbf{T}raffic \textbf{B}idirectional \textbf{E}ncoder \textbf{R}epresentations from \textbf{T}ransformer (\textbf{ET-BERT}), which pre-trains deep contextualized datagram-level representation from large-scale unlabeled data. The pre-trained model can be fine-tuned on a small number of task-specific labeled data and achieves state-of-the-art performance across five encrypted traffic classification tasks, remarkably pushing the F1 of ISCX-VPN-Service to 98.9\% (5.2\%$\uparrow$),  Cross-Platform (Android) to 92.5\% (5.4\%$\uparrow$), CSTNET-TLS 1.3 to 97.4\% (10.0\%$\uparrow$). 
		Notably, we provide explanation of the empirically powerful pre-training model by analyzing the randomness of ciphers. It gives us insights in understanding the boundary of classification ability over encrypted traffic. The code is available at: https://github.com/linwhitehat/ET-BERT.
	\end{abstract}

	
	
	
	\maketitle
	
	\raggedbottom
	
	\begin{figure}[th]
		\centering
		\includegraphics[width=\linewidth]{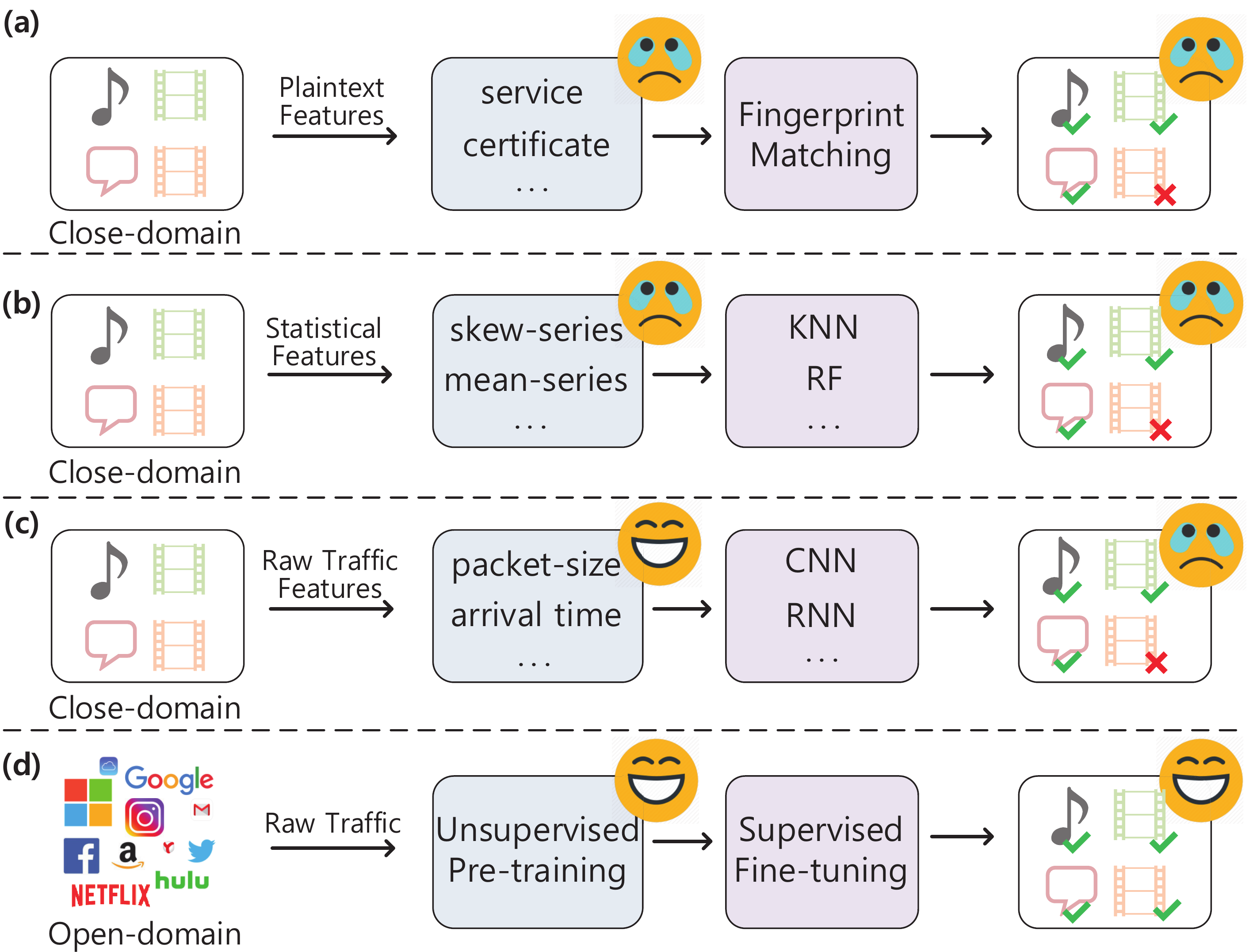}
		\caption{Four main kinds of Encrypted Traffic Classification Methods: (a) Plaintext feature based fingerprint matching. (b) Statistical feature based machine learning. (c) Raw traffic feature based ML. (d) Raw traffic based pre-training.}
		\label{fig-1}
	\end{figure}
	
	\section{INTRODUCTION}
	
	Network traffic classification, aiming to identify the category of traffic from various applications or web services, is an important technique in network management and network security \cite{BUJLOW201575, SHI201881}. Recently, traffic encryption has been widely utilized to protect the privacy and anonimity of Internet users. However, it also brings great challenges to traffic classification since the malware traffic and the cybercriminals can evade the surveillance system by privacy-enhanced encryption techniques, such as Tor, VPN, etc. Traditional methods capture patterns and keywords in the data packets from the payload, called deep packet inspection (DPI), fail to apply to the encrypted traffic. Furthermore, due to the rapid development of encryption technology, traffic classification methods for a specific kind of encrypted traffic cannot adapt well to the new environment or unseen encryption strategies \cite{Rezaei2019}. Therefore, how to capture the implicit and robust patterns in the diverse encrypted traffic and support accurate and generic traffic classification is essential to achieve high network security and effective network management.
	
	To tackle the above problem, research in encrypted traffic classification has evolved significantly over time as illustrated in Figure \ref{fig-1}. Early works \cite{Ede2020} leverage the remaining plaintext in the encrypted traffic (\textit{e.g.} certificates) to construct the fingerprint and conduct fingerprint matching for classification (Figure \ref{fig-1}(a)). However, these methods are not applicable to the newly emerging encrypted techniques (\textit{e.g.} TLS 1.3) since the plaintext becomes more sparse or obfuscated. To this end, some works \cite{Taylor2018,Panchenko2016} extract the statistical feature and employ classical machine learning algorithms to handle the encrypted traffic without plaintext (Figure \ref{fig-1}(b)). These methods highly rely on expert-designed features and have limited generalization ability. Recently, deep learning methods \cite{Liu2019,Lin2021} automatically learn complicated patterns from the raw traffic (Figure \ref{fig-1}(c)), and achieve remarkable performance improvement. However, these methods highly rely on the amount and distribution of labelled training data, which is easy to cause model bias and hard to adapt to newly emerged encryption. 
	
	In recent years, pre-training models have great breakthrough in nature language processing \cite{devlin2019}, computer vision \cite{DosovitskiyB0WZ21} and a wide range of other fields \cite{Lee2019,alsentzer-etal-2019-publicly}. Pre-training based methods adopt large unlabeled data to learn the unbiased data representations. Such data representations can be easily transferred to the downstream tasks by fine-tuning on limited amount of labeled data. In the field of encrypted traffic classification, the most recent work \cite{HeYC20} directly applies the pre-training technique and obtains obvious improvement on VPN traffic classification, but it lacks a pre-training task designed for traffic and a reasonable input representation to demonstrate the effect of the pre-training model.
	
	In this paper, we propose a novel pre-training model for classifying encrypted traffic, called \textbf{E}ncrypted \textbf{T}raffic \textbf{B}idirectional \textbf{E}ncoder \textbf{R}epresentations from \textbf{T}ransformer (ET-BERT). It aims to learn generic traffic representations from large-scale unlabeled encrypted traffic (Figure \ref{fig-1}(d)). We first propose a raw traffic representation model to transform the datagram to language-like tokens for pre-training. Each traffic flow is presented by a transmission-guided structure, denoted as BURST, which serves as the input. The proposed framework consists of two stages: pre-training and fine-tuning. Specifically, the pre-training network with Transformer structure obtains datagram-level generic traffic representations by self-supervised learning on large-scale unlabeled encrypted traffic. Thereinto, we propose two novel pre-training tasks to learn the traffic-specific patterns: the Masked BURST Model (MBM) task captures the correlated relationship between different datagram bytes in the same BURST and represent them by their context; the Same-origin BURST Prediction (SBP) task models the transmission relationships of preceding and subsequent BURST. Then, ET-BERT incorporates with the specific classification task and fine-tune the parameters with small number of task-specific labeled data.\par
	
	The main contributions of this paper are summarized as follows: 
	(1) We propose a pre-training framework for encrypted traffic classification, which leverages large-scale unlabeled encrypted traffic to learn generic datagram representation for a series of encrypted traffic classification tasks.
	(2) We newly propose two traffic-specific self-supervised pre-training tasks, \textit{e.g.} Masked BURST Model and Same-origin BURST Prediction, which capture both byte-level and BURST-level contextual relationships to obtain generic datagram representations.
	(3) ET-BERT has great generalization ability and achieves a new state-of-the-art performance over 5 encrypted traffic classification tasks, including General Encrypted Application Classification,  Encrypted Malware Classification, Encrypted Traffic Classification on VPN, Encrypted Application Classification on Tor, Encrypted Application Classification on TLS 1.3, and outperforms existing works remarkably by 5.4\%, 0.2\%, 5.2\%, 4.4\%, 10.0\%. Meanwhile, we provide theoretical explanation and analysis on the powerful performance of the pre-trained model.
	\raggedbottom
	\section{RELATED WORK}
	\label{related-work}
	\subsection{Encrypted Traffic Classification}
	\textbf{Fingerprint Construction.} Unlike the packet-inspection approach under plain-text traffic, which fails when the traffic is encrypted, some studies suggest using unencrypted protocol field information. FlowPrint \cite{Ede2020} extracts device, certificate, size, and temporal features to represent each flow and constructs a fingerprint library by clustering and cross-correlating for efficient traffic classification. However, these fingerprints are easily tampered with in virtual communication networks and lose their correct meaning, whereas our model does not rely on any plain-text information.\par
	\textbf{Statistical Methods.} Most studies of encrypted traffic exploit the statistical properties of the traffic to be independent of traffic encryption. AppScanner \cite{Taylor2018} exploits statistical features of packet size for training random forest classifiers, while BIND \cite{Naami2016} also exploits statistical features of temporality. However, it is hard to design generic statistical features to cope with the massive applications and websites that keep becoming complex, while our model does not need to rely on human-designed features.\par
	\textbf{Deep Learning Models.} Encrypted traffic classification using supervised deep learning have become a popular approach that automatically extracts discriminative features rather than relying on manual design. DF \cite{Sirinam2018} uses convolutional neural networks (CNNs) and FS-Net \cite{Liu2019} uses recurrent neural networks (RNNs) to automatically extract representations from raw packet size sequences of encrypted traffic, while Deeppacket \cite{Lotfollahi2020} and TSCRNN \cite{Lin2021} are characterizing raw payloads. However, this approach relies on a large amount of supervised data to capture valid features thus learning biased representations in imbalanced data, while our model does not rely on large labeled data.\par
	
	\subsection{Pre-training Models}
	In natural language processing, the deep bidirectional pre-training model based on Transformers achieves the best results for multiple tasks. With this representation type and structure, RoBERTa \cite{RobertA} uses dynamic masking and ALBERT \cite{Lan19} proposes sentence order prediction to improve performance by advancing unsupervised tasks. The extensions of the pre-training models include knowledge enhancement and model compression, ERNIE \cite{ZhangHLJSL19} introduces entity knowledge to improve language understanding, while DistilBERT \cite{disti} reduces the number of network layers and parameters through knowledge distillation techniques to significantly speed up model training but with a slight reduction in performance. In addition, the wide applications of pre-training models in cross-domains such as visual language as well as computer vision demonstrate their advantages of utilizing unlabeled data to help learn robust feature representations on limited labeled data.\par
	
	In encrypted traffic classification, although payloads have no semantics, Sengupta \textit{et al.} \cite{Sengupta2019} exploit the randomness difference between different ciphertexts to distinguish different applications, which suggests that the encrypted traffic is not perfectly random and implicit patterns exist. PERT \cite{HeYC20} first applies the pre-training model to migrate ALBERT to encrypted traffic classification and achieves 93.23\% performance in ISCX-VPN-Service \cite{Gerard16} on F1.
	\raggedbottom
	\begin{figure*}[th]
		\centering
		\includegraphics[width=\linewidth]{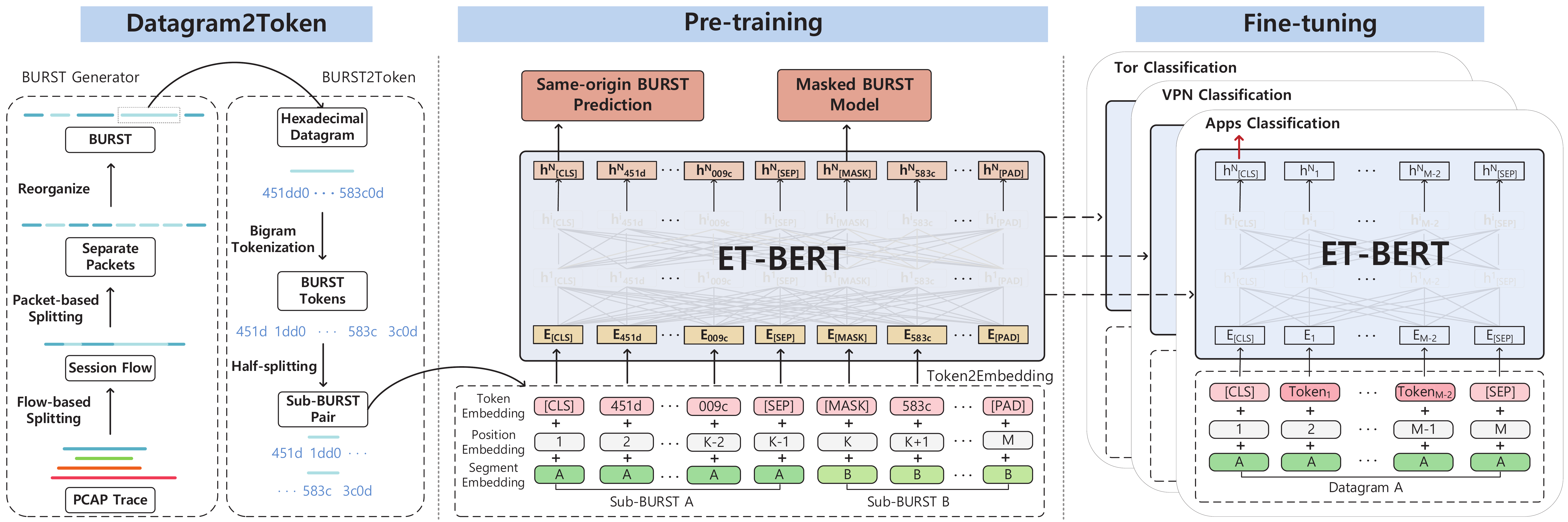}
		\caption{Overview of ET-BERT Framework.}
		\Description{.}
		\label{fig-3}
	\end{figure*}
	However, it lacks of specific design for encrypted traffic representation and the corresponding pre-training tasks, which limits its generalization ability on new encryption techniques (\textit{e.g.} TLS 1.3) according to our empirical study in Section \ref{sec5.2}. We design two pre-training tasks taking into account the structural pattern of traffic transmission and the bi-directional association of packet payloads, then use two fine-tuning strategies to better fit the traffic classification tasks.\par
	
	\section{ET-BERT}
	\label{sec4}
	
	\subsection{Model Architecture}
	\label{framework}
	In this paper, we aim to learn generic encrypted traffic representations and classify them in different scenarios (\textit{e.g.} applications, encryption protocols, or services).
	To this end, our proposed  pre-training strategy contains two main stages:  pre-training for learning generic encrypted traffic representations with large-scale unlabelled data and fine-tuning for adjusting the pre-trained model for the specific downstream task. In the pre-training stage, given the unlabelled traffic flows, the pre-trained model outputs datagram-level generic traffic representations. In the fine-tuning stage, given the target-specific labelled packets or flows, the fine-tuned model predicts its category.
	
	Encrypted traffic differs greatly from nature language and images in that it doesn’t contain human-understandable content and explicit semantic units. In order to effectively leverage the pre-training technique for encrypted traffic classification, we mainly propose three main components in ET-BERT as shown in Figure \ref{fig-3}: (1) We propose \textit{Datagram2Token} approach (Section \ref{pre-process}) to transform encrypted traffic to pattern-preserved token unit for pre-training; (2) Then two pre-training tasks, \textit{e.g.} Masked BURST Model and Same-origin BURST Prediction, are proposed to learn the contextualized datagram representations from the transition context instead of the semantic context (Section \ref{PT}); (3) To adapt to different traffic classification scenarios, we further propose two fine-tuning strategies, \textit{e.g.} packet-level fine-tuning for single packet classification and flow-level fine-tuning for single flow classification (Section \ref{FT}).
	
	The main network architecture of ET-BERT consists of multi-layer bi-directional Transformer blocks \cite{Vaswani2017}. Each of the block is composed of multi-head self-attention layers, which captures the implicit relationships between the encoded traffic units in datagrams. In this work, the network architecture consists of 12 transformer blocks with 12 attention heads in each self-attention layer. The dimension of each input token $H$ is set to 768 and the number of input tokens is 512.
	
	\subsection{Datagram2Token Traffic Representation}
	\label{pre-process}
	
	In the real network environment, huge amount of traffic contains diverse flows of different categories (\textit{e.g.} different applications, protocols or services), which makes it difficult to learn a stable and discriminative representation of a certain kind of traffic. Therefore, we first split out flows with the same IP, port and protocol from the traces before representing traffic. As a result, each splitted flow comes from the same traffic category containing a complete flow session. To further transform a flow into a word-like tokens similar to nature language, we propose a \textit{Datagram2Token} module that consists of three processes: (1) \textit{BURST Generator} extracts continues server-to-client or client-to-server packets in one session flow, named as BURST \cite{PanchenkoNZE11,Shen2021}, to represent the partial complete information of a session. (2) Then \textit{BURST2Token} process transforms the datagram in each BURST to token embeddings via the bi-gram model. Meanwhile, this process also splits a BURST into two segments preparing for the pre-training tasks. (3) Finally, \textit{Token2Emebdding} concatenates the token embedding, position embedding and segmentation embedding of each token to serve as the input representation for pre-training.
	
	\subsubsection{BURST Generator} A BURST is defined as a set of time-adjacent network packets originated from either the request or the response in a single session flow. A sequence of BURSTs characterize the pattern of network flow transmission from the application layer perspective. In the application layer, the Document Object Model (DOM) tree between web pages becomes diverse, stemming from the personalization of web services. As the client-side rendering process divides the web data into different objects (\textit{e.g.} text and images), the DOM structure generates semantic-aware fragments and subliminally affects the client's resource requests. Each generated segment forms a BURST of network, which contains a complete part of content with specific type from the DOM structure. We extract the BURSTs as input for the pre-training model.\par
	
	For the BURST, we are concerned with the source and destination of the packet. Given a trace from packet capture file as a sequence $Trace=\{flow_i, i \in \mathbb{N}^{+}\}$, where $flow = \{p_j, j \in \mathbb{N}^{+}\}$ is a session flow consisting of source-to-destination packets $p$ identified by a five-tuple (IPsrc:PORTsrc, IPdst:PORTdst, Protocol). The BURST is defined as:
	\begin{equation}\label{eq1}
		BURST = \begin{cases} B^{src} = \{p^{src}_m, m \in \mathbb{N}^{+}\}\\ B^{dst} = \{p^{dst}_n, n \in \mathbb{N}^{+}\} \end{cases}
	\end{equation}
	where $m$, $n$ denotes the maximum number of unidirectional packets of source-to-destination and destination-to-source respectively.\par
	
	\subsubsection{BURST2Token} In order to transform the BURST representation into the token representation for pre-training, we decompose the hexadecimal BURST into a sequence of units.\par
	
	To this end, we use a bi-gram to encode the hexadecimal sequence, where each unit consists of two adjacent bytes. We then use Byte-Pair Encoding for token representation, where each token unit ranges from 0 to 65535, the dictionary size $|V|$ is max expressed as 65536. In addition, we also add the special tokens \textmd{[CLS]}, \textmd{[SEP]}, \textmd{[PAD]} and \textmd{[MASK]} for training tasks. The first token of each sequence is always \textmd{[CLS]}, and the final hidden layer state associated with this token is used to represent the complete sequence for classification tasks. The token \textmd{[PAD]} is a padding notation to satisfy the minimum length requirement. The sub-BURST pair of a BURST will be separated by \textmd{[SEP]}. The token \textmd{[MASK]} appears during pre-training to learn the context of the traffic.\par
	
	As shown in Figure \ref{fig-3}, we equally divided a BURST into two sub-BURSTs for SBP task. We differentiate the sub-BURSTs by the special token \textmd{[SEP]} and the segment embedding indicating whether it belongs to segment A or segment B. We denote the segment A as sub-BURST$^{A}$ and the segment B as sub-BURST$^{B}$.\par
	
	\subsubsection{Token2Embedding}
	\label{embedding}
	We represent each token obtained in BURST2Token by three embeddings: token embedding, position embedding and segment embedding. A full token representation is constructed by summing up the aforementioned three embeddings. In this work, we take the full tokenized datagrams as original inputs. The first group of embedding vectors are randomly initialised, where the embedding dimension is $D = 768$. After N times of Transformer encoding, we obtain the final token embedding.\par
	
	\textbf{Token Embedding.} As shown in Figure \ref{fig-3}, the representation of the token learn from the lookup table in Section 3.2.2 is called token embedding $Etoken$. The final hidden vector of the input token as $Etoken \in \mathbb{R}^H$, where the embedding dimension $H$ is set to 768.\par
	
	\textbf{Position Embedding.} Since the transmission of traffic data is strongly related to the order, we use position embedding to ensure the model learn to focus on the temporal relationship of tokens by relative positions. We assign an $H$-dimensional vector to each input token for representing its position information in the sequence. We denote the position embedding as $Epos \in \mathbb{R}^H$, where the embedding dimension $H$ is set to 768.\par
	\textbf{Segment Embedding.} As mentioned in Section 3.2.2, the segment embedding of sub-BURST is denoted as $Eseg \in \mathbb{R}^H$, where the embedding dimension $H$ is set to 768. At the fine-tuning stage, we represent a packet or a flow as one segment for classification task.\par
	
	\subsection{Pre-training ET-BERT}
	\label{PT}
	Our proposed two pre-training tasks capture the contextual relationship between traffic bytes by predicting the masked token as well as the correct transmission order by predicting the Same-origin BURST. The detailed process is shown in the middle of Figure \ref{fig-3}.\par
	\textbf{Masked BURST Model.} This task is similar to the Masked Language Model utilized by BERT \cite{devlin2019}. The key difference is that traffic tokens without obvious semantics are incorporated into ET-BERT for capturing the dependencies among datagram bytes. During the pre-training, each token in the input sequence is randomly masked with 15\% probability. As the chosen token, we replace it with \textmd{[MASK]} at 80\% chance, or choose a random token to replace it or leave it unchanged at 10\% chance, respectively.\par
	
	For the masked tokens are replaced by the special token \textmd{[MASK]}, ET-BERT is trained to predict tokens at the masked positions based on the context. Benefiting from the deep bi-directional representation brought by this task, we randomly mask $k$ tokens for the input sequence $X$. We use the negative log likelihood as our loss function and formally define it as:
	\begin{equation}\label{eq10}
		L_{MBM} = -\sum_{i=1}^{k} \mathbf{log}(P(MASK_i = token_i | \bar X;\theta))
	\end{equation}
	where $\theta$ represents the set of trainable parameters of ET-BERT. The probability $P$ is modeled by the Transformer encoder with $\theta$. $\bar X$ is the representation of $X$ after masking and $MASK_i$ represents the masked token at the $i_{th}$ position in the token sequence.\par
	
	\textbf{Same-origin BURST Prediction.} The importance of BURSTs in network traffic has been declared in the previous section, and our purpose is to better learn the traffic representations by capturing the correlation of packets in BURSTs. Moreover, we consider the tight relationship between BURST structure and the web content, which is able to convey the difference between BURSTs generated from different categories of traffic. For example, there is a differentiation in the traffic by loading content separately for social networking sites with different DOM structures, \textit{e.g.} in the order of text, image, video and in the order of image, text, video. This phenomenon was also confirmed by the study \cite{WangZBKD21} for intra-domain fingerprinting.\par
	We learn the dependencies between packets inside BURST via the Same-origin BURST Prediction (SBP) task. For this task, a binary classifier is used to predict whether two sub-BURST are from the same BURST origin. Specifically, when choosing the sub-BURST$^A$ and sub-BURST$^B$ for each sub-BURST pair, 50\% of the time sub-BURST$^B$ is the actual next sub-BURST that follows sub-BURST$^A$, and 50\% of the time it is a random sub-BURST from other BURSTs. For a given input containing sub-BURST pair $B_j = (sub\mbox{-}B^A_j, sub\mbox{-}B^B_j)$ and its ground-truth label $y_j \in [0,1]$ (0 represents paired sub-BURSTs and 1 represents unpaired ones).
	\begin{equation}\label{eq11}
		L_{SBP} = -\sum_{j=1}^n \mathbf{log}(P(y_j | B_j;\theta))
	\end{equation}\par
	Overall, the final pre-training objective is the sum of the above two losses, which is defined as:
	\begin{equation}\label{eq12}
		L = L_{MBM} + L_{SBP}
	\end{equation}
	
	\textbf{Pre-training Dataset.} In this work, around 30GB of unlabeled traffic data is used for pre-training. This dataset contains two parts: (1) about 15GB traffic from the public datasets \cite{Gerard16,SharafaldinLG18}; (2) about 15GB traffic from our passively collected traffic under the China Science and Technology Network (CSTNET). Further, the dataset contains rich network protocols, such as a new encryption protocol based on UDP transport QUIC, Transport Layer Security, File Transfer Protocol, Hyper Text Transfer Protocol, Secure Shell, etc., which are common network protocols.\par
	
	\subsection{Fine-tuning ET-BERT}
	\label{FT}
	Fine-tuning can serve downstream classification tasks well because: (1) the pre-training representation is traffic class-independent and can be applied to any class of traffic representation; (2) since the input of the pre-training model is at the datagram bytes level, downstream tasks that need to classify packets and flows can be transformed into the corresponding datagram byte token to be classified by the model; (3) the special [CLS] token of the output of the pre-training model models the representation of the entire input traffic and can be employed directly for classification.\par
	Since the structure of fine-tuning and pre-training is basically identical, we input the task-specific packet or flow representations into the pre-trained ET-BERT and fine-tune all parameters in an end-to-end model. At the output layer, the \textmd{[CLS]} represenation is fed to a multi-class classifier for prediction. We propose two fine-tuning strategies to adapt the classification of different scenarios: (1) packet level as input dedicated to experimenting whether ET-BERT can adapt to more fine-grained traffic data, as \textbf{ET-BERT(packet)}; (2) flow level as input dedicated to fairly and objectively comparing ET-BERT with other methods, as \textbf{ET-BERT(flow)}. The major difference between the two fine-tuning models is the amount of information of the input traffic. We use a stitched datagram of $M$ consecutive packets in a flow as input data, where $M$ is set to 5 in our approach. The traffic data processing is described in detail in Section \ref{setup}.\par
	The cost of fine-tuning is relatively cheap compared to pre-training, and a single GPU is sufficient for a fine-tuning task.\par
	
	\section{EXPERIMENTS}
	\label{sec5}
	In this section, we conduct five encrypted traffic classification tasks (Section \ref{setup}) to prove the effectiveness of ET-BERT to solve problems of different encryption scenarios and imbalanced data distribution. We then compare our model with 11 methods (Section \ref{sec5.2}) and perform an ablation analysis of the key components of the model (Section \ref{ablation}). We further provide an interpretative analysis of the remarkable performance obtained by ET-BERT (Section \ref{interp}), and the ability to handle few-shot samples (Section \ref{small-data}).\par
	
	\begin{table}
		\small
		\caption{The Statistical Information of the Datasets.}
		\setlength\tabcolsep{2.5pt}
		\label{tab:1}
		\begin{tabular}{c|c|ccc}
			\toprule
			Task& Dataset& \#Flow& \#Packet& \#Label\\
			\midrule
			\multirow{2}{*}{GEAC} & Cross-Platform(iOS) \cite{Ede2020} & 20,858& 707,717& 196\\
			& Cross-Platform(Android) \cite{Ede2020} & 27,846& 656,044& 215\\
			\midrule
			EMC & USTC-TFC \cite{WangZZYS17} & 9,853& 97,115& 20\\
			\midrule
			\multirow{2}{*}{ETCV} & ISCX-VPN-Service \cite{Gerard16} & 3,694& 60,000& 12\\
			& ISCX-VPN-App \cite{Gerard16} & 2,329& 77,163& 17\\
			\midrule
			EACT & ISCX-Tor \cite{Arash17} & 3,021& 80,000& 16\\
			\midrule
			EAC-1.3 & CSTNET-TLS 1.3 (Ours) & 46,372& 581,709& 120\\
			\bottomrule
		\end{tabular}
	\end{table}
	
	\subsection{Experiment Setup}
	\label{setup}
	\raggedbottom
	\subsubsection{Datasets and Downstream Tasks}
	\label{dataset}
	To evaluate the effectiveness and generalization of ET-BERT, we conduct experiments across five encrypted traffic classification tasks on six public datasets and one newly proposed dataset. The tasks and the corresponding datasets are shown in Table \ref{tab:1}.\par
	
	\textbf{Task 1:} \textbf{G}eneral \textbf{E}ncrypted \textbf{A}pplication \textbf{C}lassification (\textbf{GEAC}) task aims to classify application traffic under standard encryption protocols. We test on Cross-Platform (iOS) \cite{Ede2020} and Cross-Platform (Android) \cite{Ede2020}, which contain 196 and 215 applications respectively. The iOS apps and the Android apps were collected from the top 100 Apps from the US, China and India. This dataset with the largest number of categories and long-tail data distribution over all classes.\par
	\textbf{Task 2:} \textbf{E}ncrypted \textbf{M}alware \textbf{C}lassification (\textbf{EMC}) is a collection of encrypted traffic consisting of malware and benign applications \cite{WangZZYS17}. The dataset USTC-TFC \cite{WangZZYS17} contains 10 categories of benign traffic and 10 categories of malicious traffic.\par
	\textbf{Task 3:} \textbf{E}ncrypted \textbf{T}raffic \textbf{C}lassification on \textbf{VPN} (\textbf{ETCV}) task classifies encrypted traffic that uses Virtual Private Networks (VPNs) for network communication. VPNs are popular for bypassing censorship as well as accessing geo-locked services, which is difficult to detect due to its protocol obfuscation. We use the commonly compared ISCX-VPN \cite{Gerard16}, which is constructed of 6 communication applications captured through the Canadian Institute for Cybersecurity in both VPN and non-VPN. To test ET-BERT on service and application, we further categorize the dataset by services and applications, forming the ISCX-VPN-Service dataset with 12 categories and the ISCX-VPN-App dataset with 17 applications.\par
	
	\textbf{Task 4:} \textbf{E}ncrypted \textbf{A}pplication \textbf{C}lassification on \textbf{Tor} (\textbf{EACT}) task aims to classify encrypted traffic that uses the Onion Router (Tor) for communication privacy enhancement. The dataset \cite{Arash17} is called ISCX-Tor, which contains 16 applications. This kind of traffic further obscures the behavior of the traffic by obfuscating the communication between the sender and the receiver through a distributed routing network, which is more challenging for traffic classification as the pattern extraction of traffic becomes harder.\par
	\textbf{Task 5:} \textbf{E}ncrypted \textbf{A}pplication \textbf{C}lassification on \textbf{TLS 1.3} (\textbf{EAC-1.3}) task aims to classify encrypted traffic
	\begin{table*}[th]
		\setlength\tabcolsep{3pt}
		\footnotesize
		\caption{Comparison Results on Cross-Platform, ISCX-VPN-Service and ISCX-VPN-App datasets.}
		\label{tab:4}
		\begin{tabular}{l|cccc|cccc|cccc|cccc}
			\toprule
			Dataset & \multicolumn{4}{c|}{Cross-Platform(iOS)} & \multicolumn{4}{c|}{Cross-Platform(Android)} & \multicolumn{4}{c|}{ISCX-VPN-Service} & \multicolumn{4}{c}{ISCX-VPN-App}\\
			\midrule
			Method& AC& PR& RC& F1& AC& PR& RC& F1& AC& PR& RC& F1& AC& PR& RC& F1\\
			\midrule
			AppScanner\cite{Taylor2018} & 0.3205& 0.2103& 0.2173& 0.2030& 0.3868& 0.2523& 0.2594& 0.2440 & 0.7182& 0.7339& 0.7225& 0.7197 & 0.6266& 0.4864& 0.5198& 0.4935\\
			CUMUL\cite{Panchenko2016}& 0.2910& 0.1917& 0.2081& 0.1875& 0.3525& 0.2221& 0.2409& 0.2189 & 0.5610& 0.5883& 0.5676& 0.5668 & 0.5365& 0.4129& 0.4535& 0.4236\\
			BIND\cite{Naami2016} & 0.3770& 0.2566& 0.2715& 0.2484& 0.4728& 0.3126& 0.3253& 0.3026 & 0.7534& 0.7583& 0.7488& 0.7420 & 0.6767& 0.5152& 0.5153& 0.4965\\
			K-fp\cite{Hayes2016} & 0.2155& 0.2037& 0.2069& 0.2003& 0.2248& 0.2113& 0.2104& 0.2052 & 0.6430& 0.6492& 0.6417& 0.6395& 0.6070& 0.5478& 0.5430& 0.5303\\
			FlowPrint\cite{Ede2020} & 0.9254& 0.9438& 0.9254& 0.9260& 0.8698& 0.9007& 0.8698& 0.8702 & 0.7962& 0.8042& 0.7812& 0.7820& 0.8767& 0.6697& 0.6651& 0.6531\\
			DF\cite{Sirinam2018} & 0.3106& 0.2232& 0.2179& 0.2140& 0.3862& 0.2595& 0.2620& 0.2527 & 0.7154& 0.7192& 0.7104& 0.7102& 0.6116& 0.5706& 0.4752& 0.4799\\
			FS-Net\cite{Liu2019} & 0.3712& 0.2845& 0.2754& 0.2655& 0.4846& 0.3544& 0.3365& 0.3343 & 0.7205& 0.7502& 0.7238& 0.7131& 0.6647& 0.4819& 0.4848& 0.4737\\
			GraphDApp\cite{Shen2021} & 0.3245& 0.2450& 0.2392& 0.2297& 0.4031& 0.2842& 0.2786& 0.2703 & 0.5977& 0.6045& 0.6220& 0.6036& 0.6328& 0.5900& 0.5472& 0.5558\\
			TSCRNN\cite{Lin2021} & -& -& -& -& -& -& -& -& -& 0.9270& 0.9260& 0.9260& -& -& -& -\\
			Deeppacket\cite{Lotfollahi2020} & 0.9204& 0.8963& 0.8872& 0.9034& 0.8805& 0.8004& 0.7567& 0.8138 & 0.9329& 0.9377& 0.9306& 0.9321& 0.9758& 0.9785& 0.9745& 0.9765\\
			\midrule
			PERT\cite{HeYC20} & 0.9789& 0.9621& 0.9611& 0.9584& 0.9772& 0.8628& 0.8591& 0.8550 & 0.9352& 0.9400& 0.9349& 0.9368& 0.8229& 0.7092& 0.7173& 0.6992\\
			\midrule
			ET-BERT(flow) &\textbf{0.9844}& 0.9701& 0.9632& 0.9643& \textbf{0.9865}& 0.9324& \textbf{0.9266}& \textbf{0.9246} & 0.9729& 0.9756& 0.9731& 0.9733& 0.8519& 0.7508& 0.7294& 0.7306\\
			ET-BERT(packet) & 0.9810& \textbf{0.9757}& \textbf{0.9772}& \textbf{0.9754}& 0.9728& \textbf{0.9439}& 0.9119& 0.9206 & \textbf{0.9890}& \textbf{0.9891}& \textbf{0.9890}& \textbf{0.9890}& \textbf{0.9962}& \textbf{0.9936}& \textbf{0.9938}& \textbf{0.9937}\\
			\bottomrule
		\end{tabular}
	\end{table*}
	\begin{table*}[th]
		\setlength\tabcolsep{5pt}
		\footnotesize
		\caption{Comparison Results on ISCX-Tor, USTC-TFC and CSTNET-TLS 1.3 datasets.}
		\label{tab:3}
		\begin{tabular}{l|cccc|cccc|cccc}
			\toprule
			Dataset & \multicolumn{4}{c|}{ISCX-Tor} & \multicolumn{4}{c|}{USTC-TFC} & \multicolumn{4}{c}{ CSTNET-TLS 1.3}\\
			\midrule
			Method& AC& PR& RC& F1& AC& PR& RC& F1& AC& PR& RC& F1\\
			\midrule
			AppScanner\cite{Taylor2018} & 0.6722& 0.3756& 0.4422& 0.3913& 0.8954& 0.8984& 0.8968& 0.8892 & 0.6662& 0.6246& 0.6327& 0.6201\\
			CUMUL\cite{Panchenko2016} & 0.6606& 0.3850& 0.4416& 0.3918& 0.5675& 0.6171& 0.5738& 0.5513 & 0.5391& 0.4942& 0.5060& 0.4904\\
			BIND\cite{Naami2016} & 0.7185& 0.4598& 0.4515& 0.4511& 0.8457& 0.8681& 0.8382& 0.8396 & 0.7964& 0.7605& 0.7650& 0.7560\\
			K-fp\cite{Hayes2016} & 0.6472& 0.5576& 0.5849& 0.5522& -& -& -& - & 0.4036& 0.3969& 0.4044& 0.3902\\
			FlowPrint\cite{Ede2020} & 0.9092& 0.3820& 0.3661& 0.3654& 0.8146& 0.6434& 0.7002& 0.6573 & 0.1261& 0.1354& 0.1272& 0.1116\\
			DF\cite{Sirinam2018} & 0.7533& 0.6228& 0.6010& 0.5850& 0.7787& 0.7883& 0.7819& 0.7593 & 0.7936& 0.7721& 0.7573& 0.7602\\
			FS-Net\cite{Liu2019} & 0.6071& 0.5080& 0.5350& 0.4590& 0.8846& 0.8846& 0.8920& 0.8840 & 0.8639& 0.8404& 0.8349& 0.8322\\
			GraphDApp\cite{Shen2021} & 0.6836& 0.4864& 0.4823& 0.4488& 0.8789& 0.8226& 0.8260& 0.8234 & 0.7034& 0.6464& 0.6510& 0.6440\\
			TSCRNN\cite{Lin2021} & -& 0.9490& 0.9480& 0.9480& -& 0.9870& 0.9860& 0.9870 & -& -& -& -\\
			Deeppacket\cite{Lotfollahi2020} & 0.7449& 0.7549& 0.7399& 0.7473& 0.9640& 0.9650& 0.9631& 0.9641 & 0.8019& 0.4315& 0.2689& 0.4022\\
			\midrule
			PERT\cite{HeYC20} & 0.7682& 0.4424& 0.4446& 0.4345& 0.9909& 0.9911& 0.9910& 0.9911 & 0.8915& 0.8846& 0.8719& 0.8741\\
			\midrule
			ET-BERT(flow) & 0.8311& 0.5564& 0.6448& 0.5886& \textbf{0.9929}& \textbf{0.9930}& \textbf{0.9930}& \textbf{0.9930} & 0.9510& 0.9460& 0.9419& 0.9426\\
			ET-BERT(packet) & \textbf{0.9921}& \textbf{0.9923}& \textbf{0.9921}& \textbf{0.9921}& 0.9915& 0.9915& 0.9916& 0.9916 & \textbf{0.9737}& \textbf{0.9742}& \textbf{0.9742}& \textbf{0.9741}\\
			\bottomrule
		\end{tabular}
	\end{table*}
	over new encryption protocol TLS 1.3. The dataset is our collection of 120 applications under CSTNET from March to July 2021, named as CSTNET-TLS 1.3. As we know, this is the first TLS 1.3 dataset to date. The applications are acquired from Alexa Top-5000 \cite{Alexa21} deployed with TLS 1.3 and we label each session flow by the server name indication (SNI). In CSTNET-TLS 1.3, the SNI remains accessible due to the compatibility of TLS 1.3. The ECH mechanism will disable the SNI in the future and compromise the accuracy of the labeling, but we discuss some thoughts to overcome it in Section \ref{sec6}.
	
	\textbf{Ethical Considerations.} For this research, an IRB was consulted and any identification was not utilized. Furthermore, the collection was completely passive. We have conformed to the user agreements of the corporate network where the data was collected.\par
	
	\subsubsection{Data Pre-processing.} We remove packets of Address Resolution Protocol (ARP) and Dynamic Host Configuration Protocol (DHCP), which are irrelevant to specific traffic of the transmitted content. To avoid the influence of the packet header, which may introduce biased interference in a finite set with strong identification information such as IP and port \cite{Lotfollahi2020, Li2018, Wang2020}, we removed the Ethernet header, the IP header, and protocol ports of the TCP header. In the stage of fine-tuning, we randomly select at most 500 flows and 5,000 packets from each class in all datasets. Each dataset is divided into the training set, the validation set and the testing set according to the ratio of $8:1:1$.\par
	
	\subsubsection{Evaluation Metrics and Implementation Details}
	We evaluate and compare the performance of the our model by four typical metrics, including Accuracy (AC), Precision (PR), Recall (RC), and F1 \cite{Ede2020,Zheng2020}. Macro Average \cite{LiuWWLM17} is used to avoid biased results due to imbalance between multiple categories of data by calculating the mean value of AC, PR, RC and F1 of each category. In pre-training, the batch size is 32 and the total steps is 500,000. We set the learning rate is $2 \times 10^{-5}$, and the ratio of warmup is $0.1$. We fine-tune with the AdamW optimizer for 10 epochs, where the learning rate is set to $6 \times 10^{-5}$ for flow-level, and $2 \times 10^{-5}$ for packet-level. The batch size is 32 and the dropout rate is $0.5$. All the experiments are implemented with Pytorch 1.8.0 and UER \cite{zhao2019uer}, conducted with NVIDIA Tesla V100S GPUs.\par
	
	\subsection{Comparison with State-of-the-Art Methods}
	\label{sec5.2}
	
	We compare ET-BERT with various state-of-the-art (SOTA) methods, including (1) fingerprint construction method: FlowPrint \cite{Ede2020}; (2) statistical feature methods: AppScanner \cite{Taylor2018}, CUMUL \cite{Panchenko2016}, BIND \cite{Naami2016} and k-fingerprinting (K-fp) \cite{Hayes2016}; (3) deep learning methods: Deep Fingerprinting (DF) \cite{Sirinam2018}, FS-Net \cite{Liu2019}, GraphDApp \cite{Shen2021}, TSCRNN \cite{Lin2021}, Deeppacket \cite{Lotfollahi2020}; (4) pre-training method: PERT \cite{HeYC20}.
	\begin{table*}
		\caption{Ablation Study of Key Components in ET-BERT on ISCX-VPN-App dataset.}
		\label{tab:10}
		\setlength\tabcolsep{5pt}
		\small
		\begin{tabular}{l|l|ccccccc|cccc}
			\toprule
			\multicolumn{2}{c|}{Method} & SBP& MBM& \makecell[c]{PT-P}& \makecell[c]{PT-B}& \makecell[c]{FT-f}& \makecell[c]{FT-cf}& \makecell[c]{FT-P}& AC& PR& RC& F1\\
			\midrule
			\multicolumn{2}{l|}{\textbf{ET-BERT(packet)(full model)}}& \checkmark& \checkmark& $\times$& \checkmark& $\times$& $\times$& \checkmark& \textbf{0.9471}& \textbf{0.9462}& \textbf{0.9412}& \textbf{0.9395}\\
			\midrule
			1 &\texttt{w/o SBP}& $\times$& \checkmark& $\times$& \checkmark& $\times$& $\times$& \checkmark& 0.9000& 0.9142& 0.9000& 0.8998\\
			2 &\texttt{w/o MBM}& \checkmark& $\times$& $\times$& \checkmark& $\times$& $\times$& \checkmark& 0.8471& 0.8666& 0.8471& 0.8462\\
			3 &\texttt{w/o BURST}& \checkmark& \checkmark& \checkmark& $\times$& $\times$& $\times$& \checkmark& 0.9235& 0.9386& 0.9235& 0.9258\\
			\midrule
			4 &\texttt{ET-BERT(flow)}& \checkmark& \checkmark& $\times$& \checkmark& \checkmark& $\times$& $\times$& 0.8133& 0.7661& 0.7374& 0.7387\\
			5 &\texttt{concatenated-flow(cf)}& \checkmark& \checkmark& $\times$& \checkmark& $\times$& \checkmark& $\times$& 0.8229& 0.7488& 0.6812& 0.6961\\
			\midrule
			6 &\texttt{w/o pre-training(packet)}& $\times$& $\times$& $\times$& $\times$& $\times$& $\times$& \checkmark& 0.5882& 0.6152& 0.5882& 0.5638\\
			\bottomrule
		\end{tabular}
	\end{table*}
	The experimental results are shown in Tables \ref{tab:4} and \ref{tab:3}. Additional comparison study can be found in Appendix \ref{qa}.\par
	
	\textbf{GEAC.} According to Table \ref{tab:4}, both ET-BERT(packet) and ET-BERT(flow) outperform all methods significantly. Our model obtains 1.7\% and 5.4\% respective improvement from the existing state of the art (\textit(e.g.) PERT and FlowPrint) on Cross-Platform (iOS) and Cross-Platform (Android). FlowPrint utilizes plain-text fingerprints including certificate fields to build a multi-dimensional fingerprint library for identification of applications. However, ET-BERT learns contextual relationships on ciphertext without relying on any plaintext fields. In addition, our model learns the pattern of traffic transmission structure while PERT has not been able to master.\par
	
	\textbf{EMC.} From the results on USTC-TFC as presented in Table \ref{tab:3}, we can observe that the performance of all the methods are inferior to our model.Our model achieves the best performance on F1 as 99.30\%. We observe that the malicious traffic in this dataset contain unencrypted data in application layer and this makes much easier for other models to leverage such plaintext for easier classification.\par
	\textbf{ETCV} ET-BERT achieves 5.69\% and 1.72\% improvement over the existing state-of-the-art model Deeppacket on ISCX-VPN-Service and ISCX-VPN-App. Both datasets raise the imbalance challenge, which is more severe on ISCX-VPN-App. Our model and Deeppacket alleviate the effect of imbalanced data by learning correlations between packet datagrams. Besides, ET-BERT achieves an average improvement on F1 of 25.55\% and 42.89\% for all methods except PERT, which indicates that our model has strong ability in identifying confusion traffic even in the case of imbalanced data.\par
	\textbf{EACT.} As the results on ISCX-Tor shown in Table \ref{tab:3}, ET-BERT improves 4.41\% over the existing best result obtained by TSCRNN. The initial traffic in Tor is not only multi-layer encrypted but also adversarially obfuscated. TSCRNN enriches flows by random sampling to train the model, while we exploit the intrinsic relationship of packets to achieve better classification.\par
	\textbf{EAC-1.3.} Our model is improved by 10.0\% from 87.41\% over the existing state-of-the-art as the results on CSTNET-TLS 1.3 shown in Table \ref{tab:3}. TLS 1.3 poses new challenge for the FlowPrint and Deeppacket by improving the security of the transmission and conversely. ET-BERT pushes F1 to 97.41\% by deeply representing datagrams. This indicates that the encrypted traffic datagrams over TLS 1.3 still have implicit  patterns, which is better leveraged by ET-BERT for classification.\par
	
	\subsection{Ablation Study}
	\label{ablation}
	We show ablation results to verify the contribution of each component on the widely compared ISCX-VPN-App. To fairly compare packet and flow level fine-tuning approaches, we randomly selected at most 100 packets and flows from each class as the training dataset. In Table \ref{tab:10}, PT-P and PT-B respectively represent the inputs are randomly selected adjacent packets and our proposed BURST packets in pre-training (PT). FT-f, FT-cf and FT-P respectively denote using a flow, a concatenated flow \cite{HeYC20} and a single packet in fine-tuning (FT).
	\begin{table}[th]
		\small
		\setlength\tabcolsep{2pt}
		\caption{Results of 15 Randomness Tests on 5 Ciphers.}
		\label{tab:2}
		\begin{tabular}{l|ccccc}
			\toprule
			Ciphers/Tests &AES(GCM) &AES(CBC) &CHA20 &ARC4 &3DES \\
			\midrule
			Monobit & 0.7918 & 0.2585 & 0.9761 & 0.5687 & 0.4099 \\
			Block Frequency & 0.6316 & 0.0791 & 0.0176 & 0.4821 & 0.6434 \\
			Independent Runs & 0.8824 & 0.1672 & 0.8966 & 0.7052 & 0.4241 \\
			Longest Runs & 0.7198 & 0.3148 & 0.5134 & 0.5156 & 0.2889 \\
			Spectral & 0.6202 & 0.9707 & 0.9415& 0.6729 & 0.5756 \\
			Overlapping Patterns(OP) & 0.0519 & 0.9856 & 0.1002 & 0.9089 & 0.4762 \\
			Non OP & 0.8148 & 0.1967 & 0.4445 & 0.0096 & 0.5156 \\
			Universal & 0.8501 & 0.3277 & 0.1149 & 0.0416 & 0.3062 \\
			Serial & 0.7690 & 0.4539 & 0.1600 & 0.6068 & 0.8381 \\
			Approximate Entropy & 0.9239 & 0.5226 & 0.3371 & 0.3470 & 0.3611 \\
			Cumulative Sums & 0.9496 & 0.4512 & 0.7355 & 0.1742 & 0.4043 \\
			Random Excursions(RE) & 0.1811 & 0.1232 & 0.4112 & 0.9424 & 0.9091 \\
			RE Variant & 0.4805 & 0.0119 & 0.9542 & 0.5978 & 0.9065 \\
			Matrix Rank & 0.5674 & 0.4890 & 0.0880 & 0.0504 & 0.1447 \\
			Linear Complexity & 0.6235 & 0.4519 & 0.7428 & 0.0952 & 0.9384 \\
			\bottomrule
		\end{tabular}
	\end{table}
	(1) In models '1-3', we evaluate the impact of each task and the input of pre-training. The respective decrease of 3.97\% and 9.33\% on F1 for '1' and '2' indicates that both self-supervised tasks are beneficial in providing complementary patterns for classification. In addition, we input packets instead of BURST in '3' and the F1 score decreases by 1.37\%. It proves that the BURST structure can learn the relationship between packets for better traffic classification. (2) In model '4' and '5', we evaluate the effect of fine-tuning flows in different forms. Model '4' uses consecutive packets as the input while model '5' uses packets separately as the input and concatenates the outputs at the final encoder layer, as like PERT \cite{HeYC20}. When we switch from flow to concatenated-flow, the results for model '5' drop by 4.26\%. Different packets of one flow are interdependent and our fine-tuning method for classifying flows is more beneficial.
	(3) We remove the pre-trained model to evaluate the impact of pre-training. According to model '6', we perform supervised learning on labeled data by training the Transformer model directly and the F1 score decreases remarkably by 37.57\% compared with ET-BERT.\par
	
	\subsection{Interpretability}
	\label{interp}
	
	\subsubsection{Randomness Analysis}
	
	The aforementioned results demonstrate the effectiveness and generalization of ET-BERT due to the imperfect randomness of the ciphers of encrypted payloads. An ideal encryption scheme causes the generated message to bear the maximum possible entropy \cite{Thomas2006}. However, this hypothesis is not valid in practice since different cipher implementations have varying degrees of randomness \cite{DoganaksoyEKS10}. We evaluate the strength of 5 ciphers in this paper through 15 sets of statistical tests\cite{NIST}, where \textit{p-value} = 1 indicates perfect randomness of the sequences. As shown in Table \ref{tab:2}, these ciphers indeed fail to achieve perfect randomness.\par
	\begin{figure}[t]
		\centering
		\includegraphics[width=\linewidth]{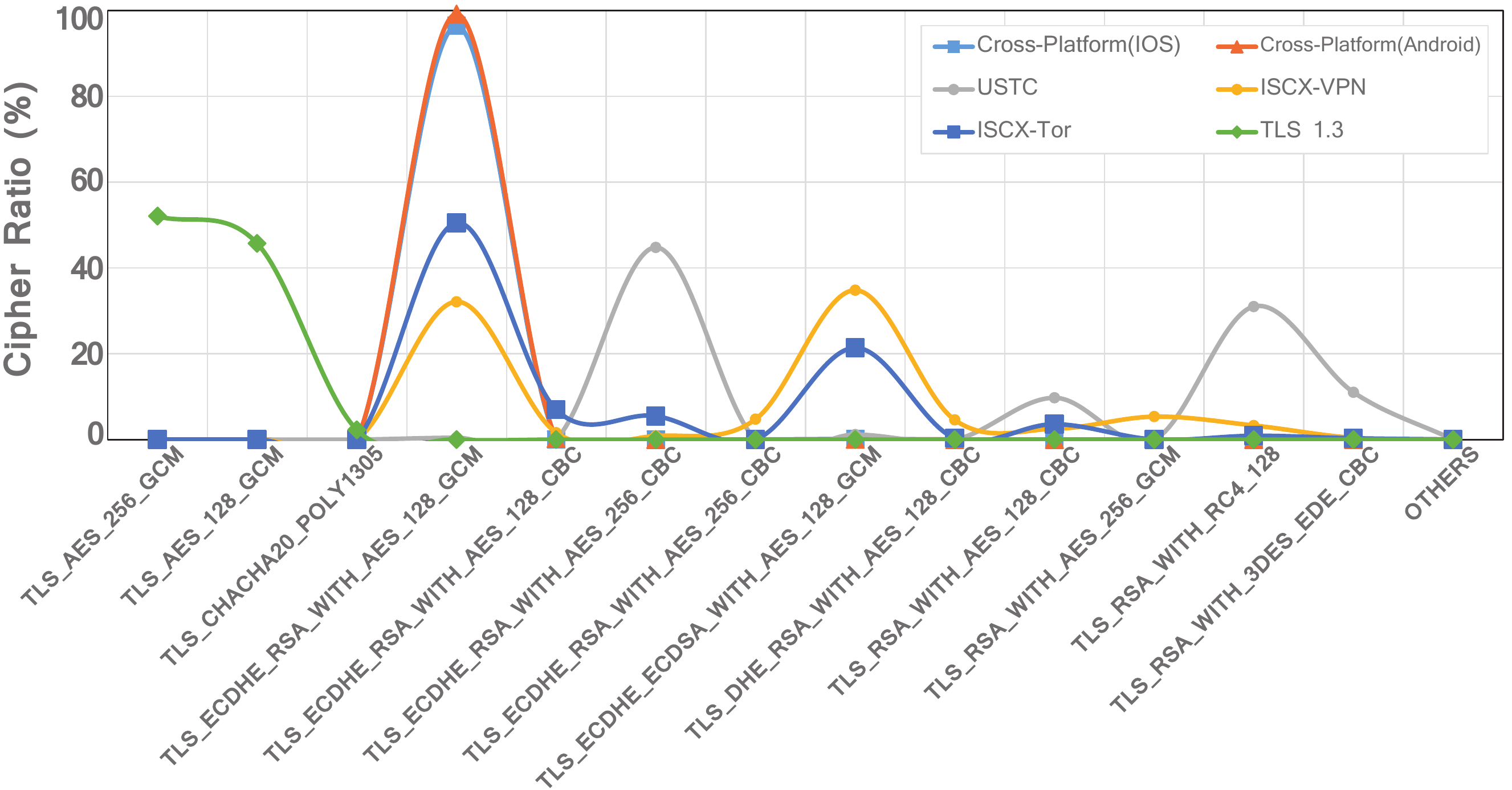}
		\caption{The Distribution of the Ciphers across all Datasets.}
		\Description{.}
		\label{fig-6}
	\end{figure}
	
	\subsubsection{Impact of Ciphers}
	
	To assess the impact of the difference ciphers of ET-BERT, we analyse the employment of ciphers for different datasets. As shown in Figure \ref{fig-6}, the horizontal coordinate indicates the ciphers that account for the top 13 types and others, and the vertical coordinate represents the percentage of each cipher. The ISCX-VPN, ISCX-Tor and USTC-TFC contain at least 3 ciphers including RC4 and 3DES with weaker randomness, while other datasets mainly consist of one cipher. According to Tables \ref{tab:4} and \ref{tab:3}, ET-BERT achieves an F1 close to 100\% on datasets with weaker randomness and the presence of greater fluctuations, average 99.14\% in ETCV, 99.21\% in EACT, and 99.30\% in EMC.\par
	
	\subsection{Few-shot Analysis}
	\label{small-data}
	
\begin{figure}[th]
		\centering
		\includegraphics[width=\linewidth]{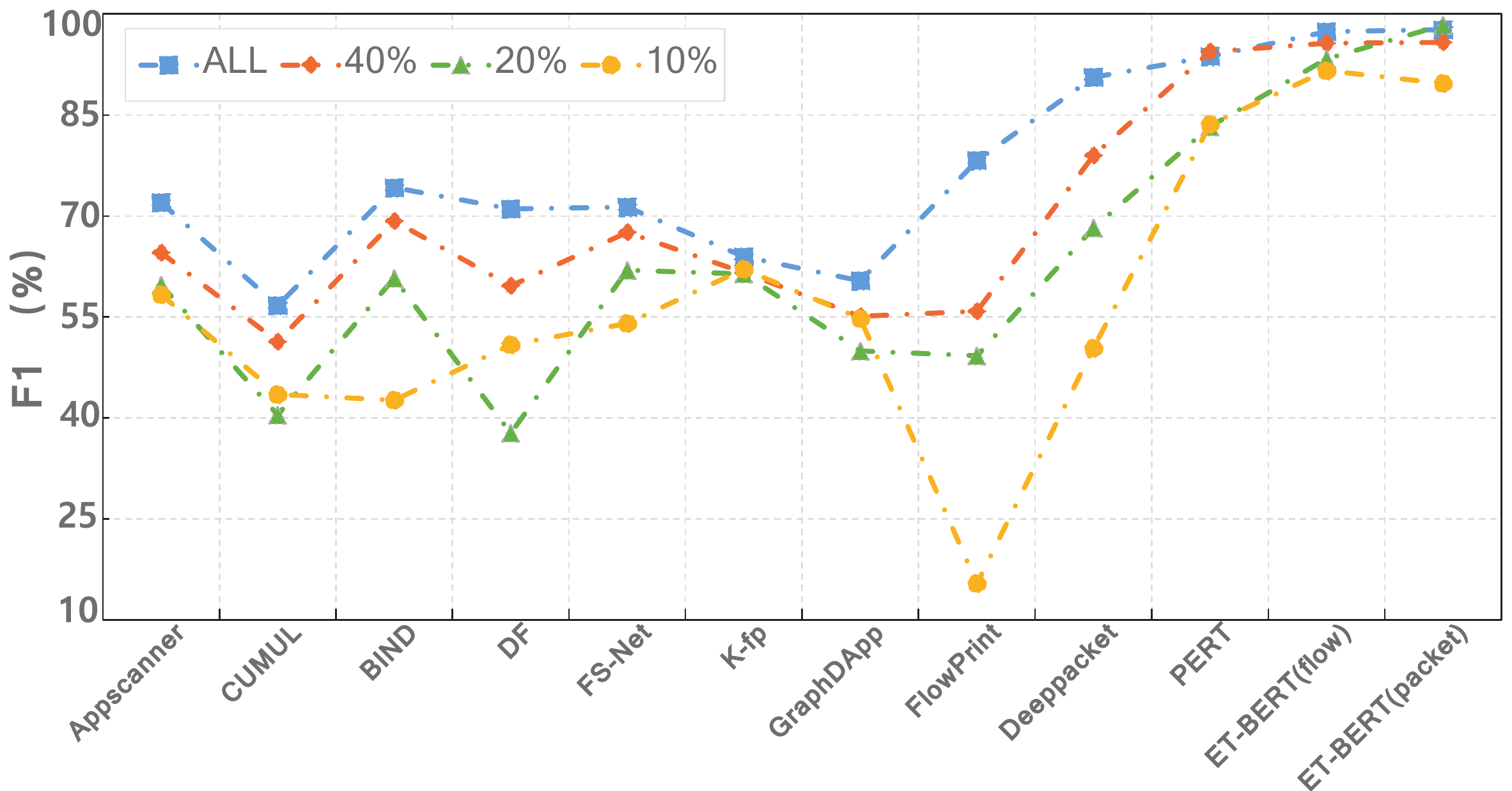}
		\caption{Comparison Results on Few-shot ISCX-VPN-App.}
		\Description{.}
		\label{fig-7}
	\end{figure}

	To validate the effectiveness and robustness of ET-BERT in few-shot settings, we design comparison experiments with different data proportions on ISCX-VPN-Service. We set the data size of each category to 500 and randomly select 40\%, 20\% and 10\% of the samples for the few-shot experiments. In Figure \ref{fig-7}, the comparison results illustrate that the pre-training method is least affected by the reduction of data size. The F1 scores of ET-BERT(packet) with 40\%, 20\%, 10\% data size are respectively 95.78\%, 98.33\% and 91.55\%. Our model achieves the best results among all methods. In contrast, traditional supervised methods (\textit{e.g.} BIND, DF, FS-Net) show substantial F1 performance degradation when the sample size is reduced, \textit{e.g.} Deeppacket's performance decreases by 40.22\% when the sample size is reduced from the full size to 10\%. This indicates that the pre-training approach solves the classification problem for the few-shot encrypted traffic more effectively.\par
	
	\section{DISCUSSION}
	\label{sec6}
	In this section, we discuss some limitations of this work, as well as the potential implications it may have in inducing further research in the field. \textbf{Generalizability:} The variability of encrypted traffic due to changes in the content of Internet services \cite{JuarezAADG14} over time will challenge the ability of our approach with fixed patterns leaned from fixed data and keeping unchanged over time.
	As the use and rise of TLS 1.3, the labeling of encrypted traffic will not be possible through SNI. We mitigate this in two ways to accommodate the ECH mechanism and thus guarantee the test of generalizability, including active visiting and labeling with unique process identifiers. \textbf{Pre-training Security:} Although ET-BERT has good robustness and generalization under a variety of encrypted traffic scenarios, it depends on the clean pre-training data. When an attacker deliberately adds low-frequency subwords as the "toxic" embeddings, a poisoned pre-trained model with a "backdoor" can be generated to force the model to predict the target class and finally fool the normally fine-tuned model on specific classification tasks \cite{KuritaMN20,KassnerS20}. However, how to construct the "toxic" tokens of encrypted traffic has not been studied yet.\par
	
	\section{CONCLUSION}
	\label{sec7}
	In this paper, we propose a new encrypted traffic representation model, ET-BERT, which can pre-train deep contextual datagram-level traffic representations from large-scale unlabeled data, then accurately classify encrypted traffic for multiple scenarios with a simple fine-tuning on a small amount of task-specific labeled data. We comprehensively evaluate the generalization and robustness of ET-BERT on 5 publicly available datasets and the TLS 1.3 dataset collected from CSTNET. ET-BERT has great generalization ability and achieves a new state-of-the-art performance over 5 encrypted traffic classification tasks, including General Encrypted Application Classification,  Encrypted Malware Classification, Encrypted Traffic Classification on VPN, Encrypted Application Classification on Tor, Encrypted Application Classification on TLS 1.3, and outperforms existing works remarkably by 5.4\%, 0.2\%, 5.2\%, 4.4\%, 10.0\%. In the future, we would like to investigate the ability of ET-BERT to predict new classes of samples and to resist sample attacks.
	
	
	\begin{acks}
		This work is supported by The National Key Research and Development Program of China No. 2021YFB3101400 and the Strategic Priority Research Program of Chinese Academy of Sciences, Grant No. XDC02040400. We are grateful to anonymous reviewers for their fruitful comments, corrections and inspiration to improve this paper. We also sincerely appreciate the shepherding from Magnus Almgren and writing help from Zhong Guan and Lulin Wang.
	\end{acks}
	
	\bibliographystyle{ACM-Reference-Format}
	\bibliography{reference}


\begin{thebibliography}{45}


\ifx \showCODEN    \undefined \def \showCODEN     #1{\unskip}     \fi
\ifx \showDOI      \undefined \def \showDOI       #1{#1}\fi
\ifx \showISBNx    \undefined \def \showISBNx     #1{\unskip}     \fi
\ifx \showISBNxiii \undefined \def \showISBNxiii  #1{\unskip}     \fi
\ifx \showISSN     \undefined \def \showISSN      #1{\unskip}     \fi
\ifx \showLCCN     \undefined \def \showLCCN      #1{\unskip}     \fi
\ifx \shownote     \undefined \def \shownote      #1{#1}          \fi
\ifx \showarticletitle \undefined \def \showarticletitle #1{#1}   \fi
\ifx \showURL      \undefined \def \showURL       {\relax}        \fi
\providecommand\bibfield[2]{#2}
\providecommand\bibinfo[2]{#2}
\providecommand\natexlab[1]{#1}
\providecommand\showeprint[2][]{arXiv:#2}

\bibitem[\protect\citeauthoryear{Al-Naami, Chandra, Mustafa, and
  et~al.}{Al-Naami et~al\mbox{.}}{2016}]%
        {Naami2016}
\bibfield{author}{\bibinfo{person}{Khaled Al-Naami}, \bibinfo{person}{Swarup
  Chandra}, \bibinfo{person}{Ahmad Mustafa}, {and} \bibinfo{person}{et al.}}
  \bibinfo{year}{2016}\natexlab{}.
\newblock \showarticletitle{Adaptive Encrypted Traffic Fingerprinting with
  Bi-Directional Dependence}. In \bibinfo{booktitle}{\emph{Proceedings of the
  32nd Annual Conference on Computer Security Applications}}
  \emph{(\bibinfo{series}{ACSAC '16})}. \bibinfo{publisher}{Association for
  Computing Machinery}, \bibinfo{pages}{177–188}.
\newblock
\showISBNx{9781450347716}


\bibitem[\protect\citeauthoryear{Alsentzer, Murphy, and et~al.}{Alsentzer
  et~al\mbox{.}}{2019}]%
        {alsentzer-etal-2019-publicly}
\bibfield{author}{\bibinfo{person}{Emily Alsentzer}, \bibinfo{person}{John
  Murphy}, {and} \bibinfo{person}{et al.}} \bibinfo{year}{2019}\natexlab{}.
\newblock \showarticletitle{Publicly Available Clinical {BERT} Embeddings}. In
  \bibinfo{booktitle}{\emph{Proceedings of the 2nd Clinical Natural Language
  Processing Workshop}}. \bibinfo{publisher}{Association for Computational
  Linguistics}, \bibinfo{address}{Minneapolis, Minnesota, USA},
  \bibinfo{pages}{72--78}.
\newblock
\urldef\tempurl%
\url{https://doi.org/10.18653/v1/W19-1909}
\showDOI{\tempurl}


\bibitem[\protect\citeauthoryear{Amazon}{Amazon}{2021}]%
        {Alexa21}
\bibfield{author}{\bibinfo{person}{Amazon}.} \bibinfo{year}{2021}\natexlab{}.
\newblock \bibinfo{title}{Alexa top sites (2021)}.
\newblock \bibinfo{howpublished}{\url{https://www.alexa.com/topsites/}}.
\newblock


\bibitem[\protect\citeauthoryear{Bujlow, Carela{-}Espa{\~{n}}ol, and
  Barlet{-}Ros}{Bujlow et~al\mbox{.}}{2015}]%
        {BUJLOW201575}
\bibfield{author}{\bibinfo{person}{Tomasz Bujlow},
  \bibinfo{person}{Valent{\'{\i}}n Carela{-}Espa{\~{n}}ol}, {and}
  \bibinfo{person}{Pere Barlet{-}Ros}.} \bibinfo{year}{2015}\natexlab{}.
\newblock \showarticletitle{Independent comparison of popular DPI tools for
  traffic classification}.
\newblock \bibinfo{journal}{\emph{Computer Networks}}  \bibinfo{volume}{76}
  (\bibinfo{date}{Jan.} \bibinfo{year}{2015}), \bibinfo{pages}{75--89}.
\newblock


\bibitem[\protect\citeauthoryear{Cover and Thomas}{Cover and Thomas}{2006}]%
        {Thomas2006}
\bibfield{author}{\bibinfo{person}{Thomas~M. Cover} {and}
  \bibinfo{person}{Joy~A. Thomas}.} \bibinfo{year}{2006}\natexlab{}.
\newblock \bibinfo{booktitle}{\emph{Elements of information theory {(2.}
  ed.)}}.
\newblock \bibinfo{publisher}{Wiley}.
\newblock


\bibitem[\protect\citeauthoryear{Devlin, Chang, Lee, and et~al.}{Devlin
  et~al\mbox{.}}{2019}]%
        {devlin2019}
\bibfield{author}{\bibinfo{person}{Jacob Devlin}, \bibinfo{person}{Ming-Wei
  Chang}, \bibinfo{person}{Kenton Lee}, {and} \bibinfo{person}{et al.}}
  \bibinfo{year}{2019}\natexlab{}.
\newblock \showarticletitle{{BERT}: Pre-training of Deep Bidirectional
  Transformers for Language Understanding}. In
  \bibinfo{booktitle}{\emph{Proceedings of the 2019 Conference of the North
  {A}merican Chapter of the Association for Computational Linguistics: Human
  Language Technologies, Volume 1 (Long and Short Papers)}}.
  \bibinfo{publisher}{Association for Computational Linguistics},
  \bibinfo{address}{Minneapolis, Minnesota}.
\newblock


\bibitem[\protect\citeauthoryear{Doganaksoy, Ege, Ko{\c{c}}ak, and
  Sulak}{Doganaksoy et~al\mbox{.}}{2010}]%
        {DoganaksoyEKS10}
\bibfield{author}{\bibinfo{person}{Ali Doganaksoy}, \bibinfo{person}{Baris
  Ege}, \bibinfo{person}{Onur Ko{\c{c}}ak}, {and} \bibinfo{person}{Fatih
  Sulak}.} \bibinfo{year}{2010}\natexlab{}.
\newblock \showarticletitle{Cryptographic Randomness Testing of Block Ciphers
  and Hash Functions}.
\newblock \bibinfo{journal}{\emph{{IACR} Cryptol. ePrint Arch.}}
  (\bibinfo{year}{2010}), \bibinfo{pages}{564}.
\newblock


\bibitem[\protect\citeauthoryear{Dosovitskiy, Beyer, and et~al.}{Dosovitskiy
  et~al\mbox{.}}{2021}]%
        {DosovitskiyB0WZ21}
\bibfield{author}{\bibinfo{person}{Alexey Dosovitskiy}, \bibinfo{person}{Lucas
  Beyer}, {and} \bibinfo{person}{et al.}} \bibinfo{year}{2021}\natexlab{}.
\newblock \showarticletitle{An Image is Worth 16x16 Words: Transformers for
  Image Recognition at Scale}. In \bibinfo{booktitle}{\emph{9th International
  Conference on Learning Representations, {ICLR} 2021, Virtual Event, Austria,
  May 3-7, 2021}}. \bibinfo{publisher}{OpenReview.net}.
\newblock
\urldef\tempurl%
\url{https://openreview.net/forum?id=YicbFdNTTy}
\showURL{%
\tempurl}


\bibitem[\protect\citeauthoryear{Draper{-}Gil., Lashkari., Mamun., and {A.
  Ghorbani}.}{Draper{-}Gil. et~al\mbox{.}}{2016}]%
        {Gerard16}
\bibfield{author}{\bibinfo{person}{Gerard Draper{-}Gil.},
  \bibinfo{person}{Arash~Habibi Lashkari.}, \bibinfo{person}{Mohammad
  Saiful~Islam Mamun.}, {and} \bibinfo{person}{Ali {A. Ghorbani}.}}
  \bibinfo{year}{2016}\natexlab{}.
\newblock \showarticletitle{Characterization of Encrypted and VPN Traffic using
  Time-related Features}. In \bibinfo{booktitle}{\emph{Proceedings of the 2nd
  International Conference on Information Systems Security and Privacy -
  ICISSP,}}. INSTICC, \bibinfo{publisher}{SciTePress},
  \bibinfo{pages}{407--414}.
\newblock


\bibitem[\protect\citeauthoryear{{Habibi Lashkari}., {Draper Gil}., Mamun., and
  Ghorbani.}{{Habibi Lashkari}. et~al\mbox{.}}{2017}]%
        {Arash17}
\bibfield{author}{\bibinfo{person}{Arash {Habibi Lashkari}.},
  \bibinfo{person}{Gerard {Draper Gil}.}, \bibinfo{person}{Mohammad
  Saiful~Islam Mamun.}, {and} \bibinfo{person}{Ali~A. Ghorbani.}}
  \bibinfo{year}{2017}\natexlab{}.
\newblock \showarticletitle{Characterization of Tor Traffic using Time based
  Features}. In \bibinfo{booktitle}{\emph{Proceedings of the 3rd International
  Conference on Information Systems Security and Privacy - ICISSP,}}. INSTICC,
  \bibinfo{publisher}{SciTePress}, \bibinfo{pages}{253--262}.
\newblock


\bibitem[\protect\citeauthoryear{harvardnlp}{harvardnlp}{2018}]%
        {tf-anota}
\bibfield{author}{\bibinfo{person}{harvardnlp}.}
  \bibinfo{year}{2018}\natexlab{}.
\newblock \bibinfo{title}{The Annotated Transformer}.
\newblock \bibinfo{howpublished}{\url
  {http://nlp.seas.harvard.edu/2018/04/03/attention.html}}.
\newblock


\bibitem[\protect\citeauthoryear{Hayes and Danezis}{Hayes and Danezis}{2016}]%
        {Hayes2016}
\bibfield{author}{\bibinfo{person}{Jamie Hayes} {and} \bibinfo{person}{George
  Danezis}.} \bibinfo{year}{2016}\natexlab{}.
\newblock \showarticletitle{k-fingerprinting: A Robust Scalable Website
  Fingerprinting Technique}. In \bibinfo{booktitle}{\emph{25th {USENIX}
  Security Symposium ({USENIX} Security 16)}}. \bibinfo{publisher}{{USENIX}
  Association}, \bibinfo{pages}{1187--1203}.
\newblock
\showISBNx{978-1-931971-32-4}


\bibitem[\protect\citeauthoryear{He, Yang, and Chen}{He et~al\mbox{.}}{2020}]%
        {HeYC20}
\bibfield{author}{\bibinfo{person}{Hong~Ye He}, \bibinfo{person}{Zhi~Guo Yang},
  {and} \bibinfo{person}{Xiang~Ning Chen}.} \bibinfo{year}{2020}\natexlab{}.
\newblock \showarticletitle{{PERT:} Payload Encoding Representation from
  Transformer for Encrypted Traffic Classification}. In
  \bibinfo{booktitle}{\emph{2020 {ITU} Kaleidoscope: Industry-Driven Digital
  Transformation, Kaleidoscope, Ha Noi, Vietnam, December 7-11, 2020}}.
  \bibinfo{publisher}{{IEEE}}, \bibinfo{pages}{1--8}.
\newblock


\bibitem[\protect\citeauthoryear{Ju{\'{a}}rez, Afroz, Acar, D{\'{\i}}az, and
  Greenstadt}{Ju{\'{a}}rez et~al\mbox{.}}{2014}]%
        {JuarezAADG14}
\bibfield{author}{\bibinfo{person}{Marc Ju{\'{a}}rez}, \bibinfo{person}{Sadia
  Afroz}, \bibinfo{person}{Gunes Acar}, \bibinfo{person}{Claudia D{\'{\i}}az},
  {and} \bibinfo{person}{Rachel Greenstadt}.} \bibinfo{year}{2014}\natexlab{}.
\newblock \showarticletitle{A Critical Evaluation of Website Fingerprinting
  Attacks}. In \bibinfo{booktitle}{\emph{Proceedings of the 2014 {ACM} {SIGSAC}
  Conference on Computer and Communications Security, Scottsdale, AZ, USA,
  November 3-7, 2014}}. \bibinfo{publisher}{{ACM}}, \bibinfo{pages}{263--274}.
\newblock


\bibitem[\protect\citeauthoryear{Kassner and Sch{\"{u}}tze}{Kassner and
  Sch{\"{u}}tze}{2020}]%
        {KassnerS20}
\bibfield{author}{\bibinfo{person}{Nora Kassner} {and} \bibinfo{person}{Hinrich
  Sch{\"{u}}tze}.} \bibinfo{year}{2020}\natexlab{}.
\newblock \showarticletitle{Negated and Misprimed Probes for Pretrained
  Language Models: Birds Can Talk, But Cannot Fly}. In
  \bibinfo{booktitle}{\emph{Proceedings of the 58th Annual Meeting of the
  Association for Computational Linguistics, {ACL} 2020, Online, July 5-10,
  2020}}. \bibinfo{publisher}{Association for Computational Linguistics},
  \bibinfo{pages}{7811--7818}.
\newblock


\bibitem[\protect\citeauthoryear{Kurita, Michel, and Neubig}{Kurita
  et~al\mbox{.}}{2020}]%
        {KuritaMN20}
\bibfield{author}{\bibinfo{person}{Keita Kurita}, \bibinfo{person}{Paul
  Michel}, {and} \bibinfo{person}{Graham Neubig}.}
  \bibinfo{year}{2020}\natexlab{}.
\newblock \showarticletitle{Weight Poisoning Attacks on Pretrained Models}. In
  \bibinfo{booktitle}{\emph{Proceedings of the 58th Annual Meeting of the
  Association for Computational Linguistics, {ACL} 2020, Online, July 5-10,
  2020}}. \bibinfo{publisher}{Association for Computational Linguistics},
  \bibinfo{pages}{2793--2806}.
\newblock


\bibitem[\protect\citeauthoryear{Lan, Chen, Goodman, and et~al.}{Lan
  et~al\mbox{.}}{2020}]%
        {Lan19}
\bibfield{author}{\bibinfo{person}{Zhenzhong Lan}, \bibinfo{person}{Mingda
  Chen}, \bibinfo{person}{Sebastian Goodman}, {and} \bibinfo{person}{et al.}}
  \bibinfo{year}{2020}\natexlab{}.
\newblock \showarticletitle{{ALBERT:} {A} Lite {BERT} for Self-supervised
  Learning of Language Representations}. \bibinfo{publisher}{{ICLR}}.
\newblock


\bibitem[\protect\citeauthoryear{Lee, Yoon, and et~al.}{Lee
  et~al\mbox{.}}{2020}]%
        {LeeYKKKSK20}
\bibfield{author}{\bibinfo{person}{Jinhyuk Lee}, \bibinfo{person}{Wonjin Yoon},
  {and} \bibinfo{person}{et al.}} \bibinfo{year}{2020}\natexlab{}.
\newblock \showarticletitle{BioBERT: a pre-trained biomedical language
  representation model for biomedical text mining}.
\newblock \bibinfo{journal}{\emph{Bioinform.}} \bibinfo{volume}{36},
  \bibinfo{number}{4} (\bibinfo{year}{2020}), \bibinfo{pages}{1234--1240}.
\newblock
\urldef\tempurl%
\url{https://doi.org/10.1093/bioinformatics/btz682}
\showDOI{\tempurl}


\bibitem[\protect\citeauthoryear{Li, Xiao, Ni, and et~al.}{Li
  et~al\mbox{.}}{2018}]%
        {Li2018}
\bibfield{author}{\bibinfo{person}{Rui Li}, \bibinfo{person}{Xi Xiao},
  \bibinfo{person}{Shiguang Ni}, {and} \bibinfo{person}{et al.}}
  \bibinfo{year}{2018}\natexlab{}.
\newblock \showarticletitle{Byte Segment Neural Network for Network Traffic
  Classification}. In \bibinfo{booktitle}{\emph{26th {IEEE/ACM} International
  Symposium on Quality of Service, IWQoS 2018, Banff, AB, Canada, June 4-6,
  2018}}. \bibinfo{pages}{1--10}.
\newblock


\bibitem[\protect\citeauthoryear{Li, Choi, and et~al.}{Li
  et~al\mbox{.}}{2022}]%
        {alphacode}
\bibfield{author}{\bibinfo{person}{Yujia Li}, \bibinfo{person}{David Choi},
  {and} \bibinfo{person}{et al.}} \bibinfo{year}{2022}\natexlab{}.
\newblock \bibinfo{title}{Competition-Level Code Generation with AlphaCode}.
\newblock
\newblock


\bibitem[\protect\citeauthoryear{Lin, Xu, and Gao}{Lin et~al\mbox{.}}{2021}]%
        {Lin2021}
\bibfield{author}{\bibinfo{person}{Kunda Lin}, \bibinfo{person}{Xiaolong Xu},
  {and} \bibinfo{person}{Honghao Gao}.} \bibinfo{year}{2021}\natexlab{}.
\newblock \showarticletitle{{TSCRNN:} {A} novel classification scheme of
  encrypted traffic based on flow spatiotemporal features for efficient
  management of IIoT}.
\newblock \bibinfo{journal}{\emph{Comput. Networks}}  \bibinfo{volume}{190}
  (\bibinfo{year}{2021}), \bibinfo{pages}{107974}.
\newblock


\bibitem[\protect\citeauthoryear{Liu, He, Xiong, Cao, and Li}{Liu
  et~al\mbox{.}}{2019a}]%
        {Liu2019}
\bibfield{author}{\bibinfo{person}{Chang Liu}, \bibinfo{person}{Longtao He},
  \bibinfo{person}{Gang Xiong}, \bibinfo{person}{Zigang Cao}, {and}
  \bibinfo{person}{Zhen Li}.} \bibinfo{year}{2019}\natexlab{a}.
\newblock \showarticletitle{FS-Net: {A} Flow Sequence Network For Encrypted
  Traffic Classification}. In \bibinfo{booktitle}{\emph{2019 {IEEE} Conference
  on Computer Communications, {INFOCOM} 2019, Paris, France, April 29 - May 2,
  2019}}. \bibinfo{publisher}{{IEEE}}, \bibinfo{pages}{1171--1179}.
\newblock


\bibitem[\protect\citeauthoryear{Liu, Wang, Wang, Lv, and Konan}{Liu
  et~al\mbox{.}}{2017}]%
        {LiuWWLM17}
\bibfield{author}{\bibinfo{person}{Chuan Liu}, \bibinfo{person}{Wenyong Wang},
  \bibinfo{person}{Meng Wang}, \bibinfo{person}{Fengmao Lv}, {and}
  \bibinfo{person}{Martin Konan}.} \bibinfo{year}{2017}\natexlab{}.
\newblock \showarticletitle{An efficient instance selection algorithm to
  reconstruct training set for support vector machine}.
\newblock \bibinfo{journal}{\emph{Knowl. Based Syst.}}  \bibinfo{volume}{116}
  (\bibinfo{year}{2017}), \bibinfo{pages}{58--73}.
\newblock


\bibitem[\protect\citeauthoryear{Liu, Ott, Goyal, Du, Joshi, Chen, Levy, Lewis,
  Zettlemoyer, and Stoyanov}{Liu et~al\mbox{.}}{2019b}]%
        {RobertA}
\bibfield{author}{\bibinfo{person}{Yinhan Liu}, \bibinfo{person}{Myle Ott},
  \bibinfo{person}{Naman Goyal}, \bibinfo{person}{Jingfei Du},
  \bibinfo{person}{Mandar Joshi}, \bibinfo{person}{Danqi Chen},
  \bibinfo{person}{Omer Levy}, \bibinfo{person}{Mike Lewis},
  \bibinfo{person}{Luke Zettlemoyer}, {and} \bibinfo{person}{Veselin
  Stoyanov}.} \bibinfo{year}{2019}\natexlab{b}.
\newblock \showarticletitle{RoBERTa: {A} Robustly Optimized {BERT} Pretraining
  Approach}.
\newblock \bibinfo{journal}{\emph{CoRR}}  \bibinfo{volume}{abs/1907.11692}
  (\bibinfo{year}{2019}).
\newblock


\bibitem[\protect\citeauthoryear{Lotfollahi, Siavoshani, Zade, and
  Saberian}{Lotfollahi et~al\mbox{.}}{2020}]%
        {Lotfollahi2020}
\bibfield{author}{\bibinfo{person}{Mohammad Lotfollahi},
  \bibinfo{person}{Mahdi~Jafari Siavoshani}, \bibinfo{person}{Ramin
  Shirali~Hossein Zade}, {and} \bibinfo{person}{Mohammdsadegh Saberian}.}
  \bibinfo{year}{2020}\natexlab{}.
\newblock \showarticletitle{Deep packet: a novel approach for encrypted traffic
  classification using deep learning}.
\newblock \bibinfo{journal}{\emph{Soft Comput}} \bibinfo{volume}{24},
  \bibinfo{number}{3} (\bibinfo{year}{2020}), \bibinfo{pages}{1999--2012}.
\newblock


\bibitem[\protect\citeauthoryear{of~Standards and Technology}{of~Standards and
  Technology}{[n.d.]}]%
        {NIST}
\bibfield{author}{\bibinfo{person}{National~Institute of Standards} {and}
  \bibinfo{person}{Technology}.} \bibinfo{year}{[n.d.]}\natexlab{}.
\newblock \bibinfo{title}{Random Bit Generation: Guide to the Statistical
  Tests}.
\newblock \bibinfo{howpublished}{\url
  {https://csrc.nist.gov/Projects/Random-Bit-Generation/Documentation-and-Software/Guide-to-the-Statistical-Tests}}.
\newblock


\bibitem[\protect\citeauthoryear{Panchenko, Lanze, Pennekamp, and
  et~al.}{Panchenko et~al\mbox{.}}{2016}]%
        {Panchenko2016}
\bibfield{author}{\bibinfo{person}{Andriy Panchenko}, \bibinfo{person}{Fabian
  Lanze}, \bibinfo{person}{Jan Pennekamp}, {and} \bibinfo{person}{et al.}}
  \bibinfo{year}{2016}\natexlab{}.
\newblock \showarticletitle{Website Fingerprinting at Internet Scale}. In
  \bibinfo{booktitle}{\emph{23rd Annual Network and Distributed System Security
  Symposium, {NDSS} 2016, San Diego, California, USA, February 21-24, 2016}}.
  \bibinfo{publisher}{The Internet Society}.
\newblock


\bibitem[\protect\citeauthoryear{Panchenko, Niessen, Zinnen, and
  Engel}{Panchenko et~al\mbox{.}}{2011}]%
        {PanchenkoNZE11}
\bibfield{author}{\bibinfo{person}{Andriy Panchenko}, \bibinfo{person}{Lukas
  Niessen}, \bibinfo{person}{Andreas Zinnen}, {and} \bibinfo{person}{Thomas
  Engel}.} \bibinfo{year}{2011}\natexlab{}.
\newblock \showarticletitle{Website fingerprinting in onion routing based
  anonymization networks}. In \bibinfo{booktitle}{\emph{Proceedings of the 10th
  annual {ACM} workshop on Privacy in the electronic society, {WPES} 2011,
  Chicago, IL, USA, October 17, 2011}}. \bibinfo{publisher}{{ACM}},
  \bibinfo{pages}{103--114}.
\newblock


\bibitem[\protect\citeauthoryear{Rezaei and Liu}{Rezaei and Liu}{2019}]%
        {Rezaei2019}
\bibfield{author}{\bibinfo{person}{Shahbaz Rezaei} {and} \bibinfo{person}{Xin
  Liu}.} \bibinfo{year}{2019}\natexlab{}.
\newblock \showarticletitle{Deep Learning for Encrypted Traffic Classification:
  An Overview}.
\newblock \bibinfo{journal}{\emph{{IEEE} Commun. Mag.}} \bibinfo{volume}{57},
  \bibinfo{number}{5} (\bibinfo{year}{2019}), \bibinfo{pages}{76--81}.
\newblock


\bibitem[\protect\citeauthoryear{Sanh, Debut, Chaumond, and Wolf}{Sanh
  et~al\mbox{.}}{2019}]%
        {disti}
\bibfield{author}{\bibinfo{person}{Victor Sanh}, \bibinfo{person}{Lysandre
  Debut}, \bibinfo{person}{Julien Chaumond}, {and} \bibinfo{person}{Thomas
  Wolf}.} \bibinfo{year}{2019}\natexlab{}.
\newblock \showarticletitle{DistilBERT, a distilled version of {BERT:} smaller,
  faster, cheaper and lighter}.
\newblock \bibinfo{journal}{\emph{CoRR}}  \bibinfo{volume}{abs/1910.01108}
  (\bibinfo{year}{2019}).
\newblock


\bibitem[\protect\citeauthoryear{Sengupta, Ganguly, De, and
  Chakraborty}{Sengupta et~al\mbox{.}}{2019}]%
        {Sengupta2019}
\bibfield{author}{\bibinfo{person}{Satadal Sengupta}, \bibinfo{person}{Niloy
  Ganguly}, \bibinfo{person}{Pradipta De}, {and} \bibinfo{person}{Sandip
  Chakraborty}.} \bibinfo{year}{2019}\natexlab{}.
\newblock \showarticletitle{Exploiting Diversity in Android {TLS}
  Implementations for Mobile App Traffic Classification}. In
  \bibinfo{booktitle}{\emph{The World Wide Web Conference, {WWW} 2019, San
  Francisco, CA, USA, May 13-17, 2019}}. \bibinfo{publisher}{{ACM}},
  \bibinfo{pages}{1657--1668}.
\newblock


\bibitem[\protect\citeauthoryear{Sharafaldin, Lashkari, and
  Ghorbani}{Sharafaldin et~al\mbox{.}}{2018}]%
        {SharafaldinLG18}
\bibfield{author}{\bibinfo{person}{Iman Sharafaldin},
  \bibinfo{person}{Arash~Habibi Lashkari}, {and} \bibinfo{person}{Ali~A.
  Ghorbani}.} \bibinfo{year}{2018}\natexlab{}.
\newblock \showarticletitle{Toward Generating a New Intrusion Detection Dataset
  and Intrusion Traffic Characterization}. In
  \bibinfo{booktitle}{\emph{Proceedings of the 4th International Conference on
  Information Systems Security and Privacy, {ICISSP} 2018, Funchal, Madeira -
  Portugal, January 22-24, 2018}}. \bibinfo{publisher}{SciTePress},
  \bibinfo{pages}{108--116}.
\newblock


\bibitem[\protect\citeauthoryear{Shen, Zhang, Zhu, and et~al.}{Shen
  et~al\mbox{.}}{2021}]%
        {Shen2021}
\bibfield{author}{\bibinfo{person}{Meng Shen}, \bibinfo{person}{Jinpeng Zhang},
  \bibinfo{person}{Liehuang Zhu}, {and} \bibinfo{person}{et al.}}
  \bibinfo{year}{2021}\natexlab{}.
\newblock \showarticletitle{Accurate Decentralized Application Identification
  via Encrypted Traffic Analysis Using Graph Neural Networks}.
\newblock \bibinfo{journal}{\emph{{IEEE} Trans. Inf. Forensics Secur.}}
  \bibinfo{volume}{16} (\bibinfo{year}{2021}), \bibinfo{pages}{2367--2380}.
\newblock


\bibitem[\protect\citeauthoryear{Shi, Li, Zhang, and et~al.}{Shi
  et~al\mbox{.}}{2018}]%
        {SHI201881}
\bibfield{author}{\bibinfo{person}{Hongtao Shi}, \bibinfo{person}{Hongping Li},
  \bibinfo{person}{Dan Zhang}, {and} \bibinfo{person}{et al.}}
  \bibinfo{year}{2018}\natexlab{}.
\newblock \showarticletitle{An efficient feature generation approach based on
  deep learning and feature selection techniques for traffic classification}.
\newblock \bibinfo{journal}{\emph{Computer Networks}}  \bibinfo{volume}{132}
  (\bibinfo{date}{Feb.} \bibinfo{year}{2018}), \bibinfo{pages}{81--98}.
\newblock


\bibitem[\protect\citeauthoryear{Sirinam, Imani, Ju{\'{a}}rez, and
  et~al.}{Sirinam et~al\mbox{.}}{2018}]%
        {Sirinam2018}
\bibfield{author}{\bibinfo{person}{Payap Sirinam}, \bibinfo{person}{Mohsen
  Imani}, \bibinfo{person}{Marc Ju{\'{a}}rez}, {and} \bibinfo{person}{et al.}}
  \bibinfo{year}{2018}\natexlab{}.
\newblock \showarticletitle{Deep Fingerprinting: Undermining Website
  Fingerprinting Defenses with Deep Learning}. In
  \bibinfo{booktitle}{\emph{Proceedings of the 2018 {ACM} {SIGSAC} Conference
  on Computer and Communications Security, {CCS} 2018, Toronto, ON, Canada,
  October 15-19}}. \bibinfo{pages}{1928--1943}.
\newblock


\bibitem[\protect\citeauthoryear{Taylor, Spolaor, Conti, and Martinovic}{Taylor
  et~al\mbox{.}}{2018}]%
        {Taylor2018}
\bibfield{author}{\bibinfo{person}{Vincent~F. Taylor},
  \bibinfo{person}{Riccardo Spolaor}, \bibinfo{person}{Mauro Conti}, {and}
  \bibinfo{person}{Ivan Martinovic}.} \bibinfo{year}{2018}\natexlab{}.
\newblock \showarticletitle{Robust Smartphone App Identification via Encrypted
  Network Traffic Analysis}.
\newblock \bibinfo{journal}{\emph{IEEE Transactions on Information Forensics
  and Security}} \bibinfo{volume}{13}, \bibinfo{number}{1}
  (\bibinfo{date}{Jan.} \bibinfo{year}{2018}), \bibinfo{pages}{63--78}.
\newblock


\bibitem[\protect\citeauthoryear{van Ede, Bortolameotti, Continella, and
  et~al.}{van Ede et~al\mbox{.}}{2020}]%
        {Ede2020}
\bibfield{author}{\bibinfo{person}{Thijs van Ede}, \bibinfo{person}{Riccardo
  Bortolameotti}, \bibinfo{person}{Andrea Continella}, {and}
  \bibinfo{person}{et al.}} \bibinfo{year}{2020}\natexlab{}.
\newblock \showarticletitle{FlowPrint: Semi-Supervised Mobile-App
  Fingerprinting on Encrypted Network Traffic}. In
  \bibinfo{booktitle}{\emph{27th Annual Network and Distributed System Security
  Symposium, {NDSS} 2020, San Diego, California, USA, February 23-26, 2020}}.
  \bibinfo{publisher}{The Internet Society}.
\newblock


\bibitem[\protect\citeauthoryear{Vaswani, Shazeer, Parmar, Uszkoreit, and
  et~al.}{Vaswani et~al\mbox{.}}{2017}]%
        {Vaswani2017}
\bibfield{author}{\bibinfo{person}{Ashish Vaswani}, \bibinfo{person}{Noam
  Shazeer}, \bibinfo{person}{Niki Parmar}, \bibinfo{person}{Jakob Uszkoreit},
  {and} \bibinfo{person}{et al.}} \bibinfo{year}{2017}\natexlab{}.
\newblock \showarticletitle{Attention is All You Need}. In
  \bibinfo{booktitle}{\emph{Proceedings of the 31st International Conference on
  Neural Information Processing Systems}} \emph{(\bibinfo{series}{NIPS'17})}.
  \bibinfo{publisher}{Curran Associates Inc.}, \bibinfo{pages}{6000–6010}.
\newblock
\showISBNx{9781510860964}


\bibitem[\protect\citeauthoryear{Wang, Zhang, Bai, Ko, and Dong}{Wang
  et~al\mbox{.}}{2021}]%
        {WangZBKD21}
\bibfield{author}{\bibinfo{person}{Kailong Wang}, \bibinfo{person}{Junzhe
  Zhang}, \bibinfo{person}{Guangdong Bai}, \bibinfo{person}{Ryan K.~L. Ko},
  {and} \bibinfo{person}{Jin~Song Dong}.} \bibinfo{year}{2021}\natexlab{}.
\newblock \showarticletitle{It's Not Just the Site, It's the Contents:
  Intra-domain Fingerprinting Social Media Websites Through {CDN} Bursts}. In
  \bibinfo{booktitle}{\emph{{WWW} '21: The Web Conference 2021, Virtual Event /
  Ljubljana, Slovenia, April 19-23, 2021}}. \bibinfo{publisher}{{ACM} /
  {IW3C2}}, \bibinfo{pages}{2142--2153}.
\newblock


\bibitem[\protect\citeauthoryear{Wang, Li, Ye, and et~al.}{Wang
  et~al\mbox{.}}{2020}]%
        {Wang2020}
\bibfield{author}{\bibinfo{person}{Pan Wang}, \bibinfo{person}{Shuhang Li},
  \bibinfo{person}{Feng Ye}, {and} \bibinfo{person}{et al.}}
  \bibinfo{year}{2020}\natexlab{}.
\newblock \showarticletitle{PacketCGAN: Exploratory Study of Class Imbalance
  for Encrypted Traffic Classification Using {CGAN}}. In
  \bibinfo{booktitle}{\emph{2020 {IEEE} International Conference on
  Communications, {ICC} 2020, Dublin, Ireland}}. \bibinfo{pages}{1--7}.
\newblock


\bibitem[\protect\citeauthoryear{Wang, Zhu, Zeng, and et~al.}{Wang
  et~al\mbox{.}}{2017}]%
        {WangZZYS17}
\bibfield{author}{\bibinfo{person}{Wei Wang}, \bibinfo{person}{Ming Zhu},
  \bibinfo{person}{Xuewen Zeng}, {and} \bibinfo{person}{et al.}}
  \bibinfo{year}{2017}\natexlab{}.
\newblock \showarticletitle{Malware traffic classification using convolutional
  neural network for representation learning}. In
  \bibinfo{booktitle}{\emph{2017 International Conference on Information
  Networking, {ICOIN} 2017, Da Nang, Vietnam, January 11-13, 2017}}.
  \bibinfo{publisher}{{IEEE}}, \bibinfo{pages}{712--717}.
\newblock


\bibitem[\protect\citeauthoryear{Wu, Schuster, Chen, and et~al.}{Wu
  et~al\mbox{.}}{2016}]%
        {WuSCLNMKCGMKSJL16}
\bibfield{author}{\bibinfo{person}{Yonghui Wu}, \bibinfo{person}{Mike
  Schuster}, \bibinfo{person}{Zhifeng Chen}, {and} \bibinfo{person}{et al.}}
  \bibinfo{year}{2016}\natexlab{}.
\newblock \showarticletitle{Google's Neural Machine Translation System:
  Bridging the Gap between Human and Machine Translation}.
\newblock \bibinfo{howpublished}{\url{http://arxiv.org/abs/1609.08144}}.
\newblock   \bibinfo{volume}{abs/1609.08144} (\bibinfo{year}{2016}).
\newblock


\bibitem[\protect\citeauthoryear{Zhang, Han, Liu, Jiang, Sun, and Liu}{Zhang
  et~al\mbox{.}}{2019}]%
        {ZhangHLJSL19}
\bibfield{author}{\bibinfo{person}{Zhengyan Zhang}, \bibinfo{person}{Xu Han},
  \bibinfo{person}{Zhiyuan Liu}, \bibinfo{person}{Xin Jiang},
  \bibinfo{person}{Maosong Sun}, {and} \bibinfo{person}{Qun Liu}.}
  \bibinfo{year}{2019}\natexlab{}.
\newblock \showarticletitle{{ERNIE:} Enhanced Language Representation with
  Informative Entities}. In \bibinfo{booktitle}{\emph{Proceedings of the 57th
  Conference of the Association for Computational Linguistics, {ACL} 2019,
  Florence, Italy, July 28- August 2, 2019, Volume 1: Long Papers}}.
  \bibinfo{publisher}{Association for Computational Linguistics},
  \bibinfo{pages}{1441--1451}.
\newblock


\bibitem[\protect\citeauthoryear{Zhao, Chen, Zhang, and et~al.}{Zhao
  et~al\mbox{.}}{2019}]%
        {zhao2019uer}
\bibfield{author}{\bibinfo{person}{Zhe Zhao}, \bibinfo{person}{Hui Chen},
  \bibinfo{person}{Jinbin Zhang}, {and} \bibinfo{person}{et al.}}
  \bibinfo{year}{2019}\natexlab{}.
\newblock \showarticletitle{UER: An Open-Source Toolkit for Pre-training
  Models}.
\newblock \bibinfo{journal}{\emph{EMNLP-IJCNLP 2019}} (\bibinfo{year}{2019}),
  \bibinfo{pages}{241}.
\newblock


\bibitem[\protect\citeauthoryear{Zheng, Gou, Yan, and et~al.}{Zheng
  et~al\mbox{.}}{2020}]%
        {Zheng2020}
\bibfield{author}{\bibinfo{person}{Wenbo Zheng}, \bibinfo{person}{Chao Gou},
  \bibinfo{person}{Lan Yan}, {and} \bibinfo{person}{et al.}}
  \bibinfo{year}{2020}\natexlab{}.
\newblock \showarticletitle{Learning to Classify: A Flow-Based Relation Network
  for Encrypted Traffic Classification}. In
  \bibinfo{booktitle}{\emph{Proceedings of The Web Conference 2020}}.
  \bibinfo{publisher}{Association for Computing Machinery},
  \bibinfo{pages}{13--22}.
\newblock
\showISBNx{9781450370233}


\end{thebibliography}
	
	\appendix
	
	\section{Additional Comparison Study}
	\begin{figure*}[th]
		\centering
		\subfigure[Representation with ET-BERT]{
			\includegraphics[width=0.24\linewidth,height = 0.2\textwidth]{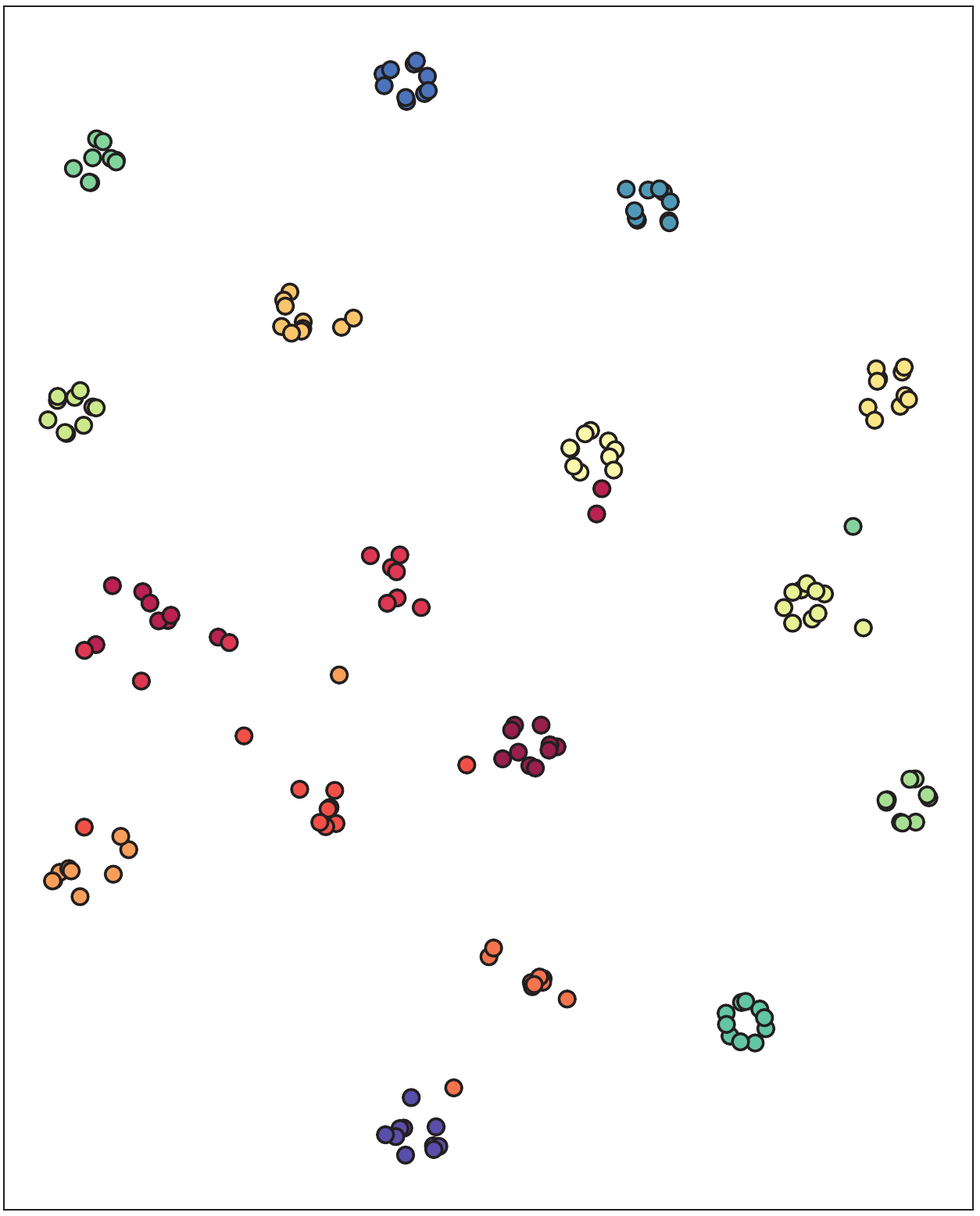}}
		\quad
		\subfigure[Representation with Transformer at packet]{
			\includegraphics[width=0.25\linewidth,height = 0.2\textwidth]{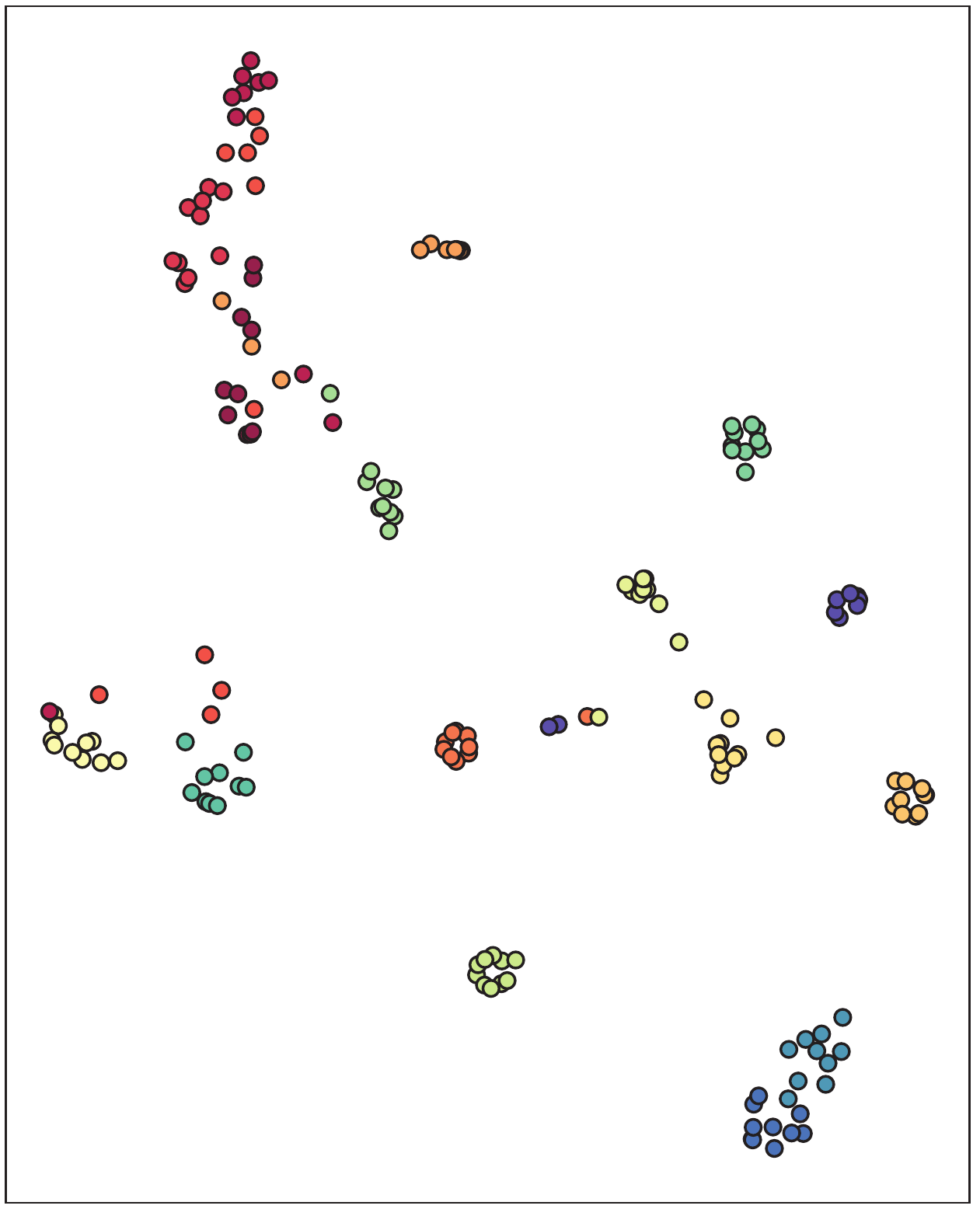}}
		\quad
		\subfigure[Representation with Deeppacket]{
			\includegraphics[width=0.24\linewidth,height = 0.2\textwidth]{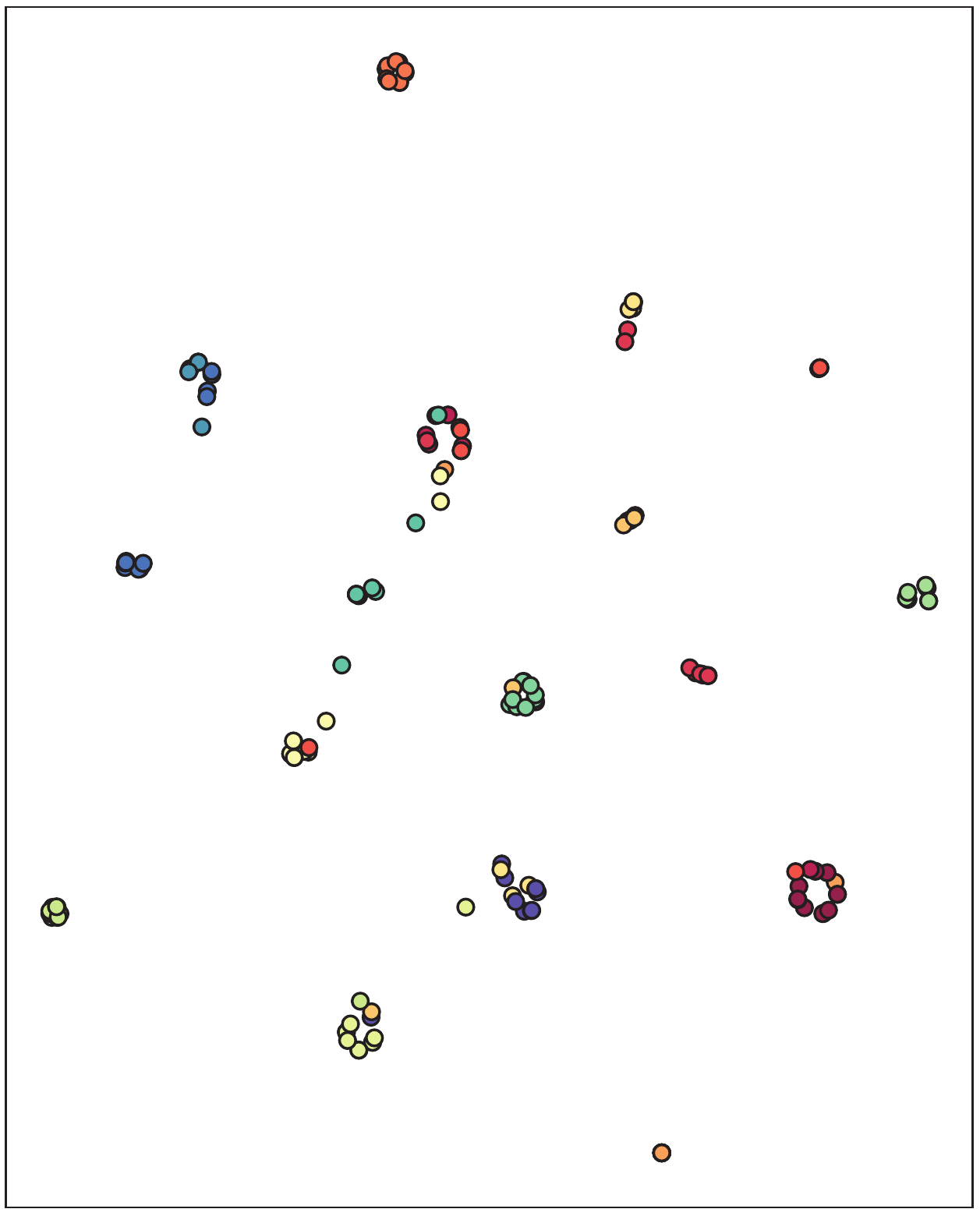}}
		\subfigure{
			\includegraphics[width=0.05\linewidth,height = 0.2\textwidth]{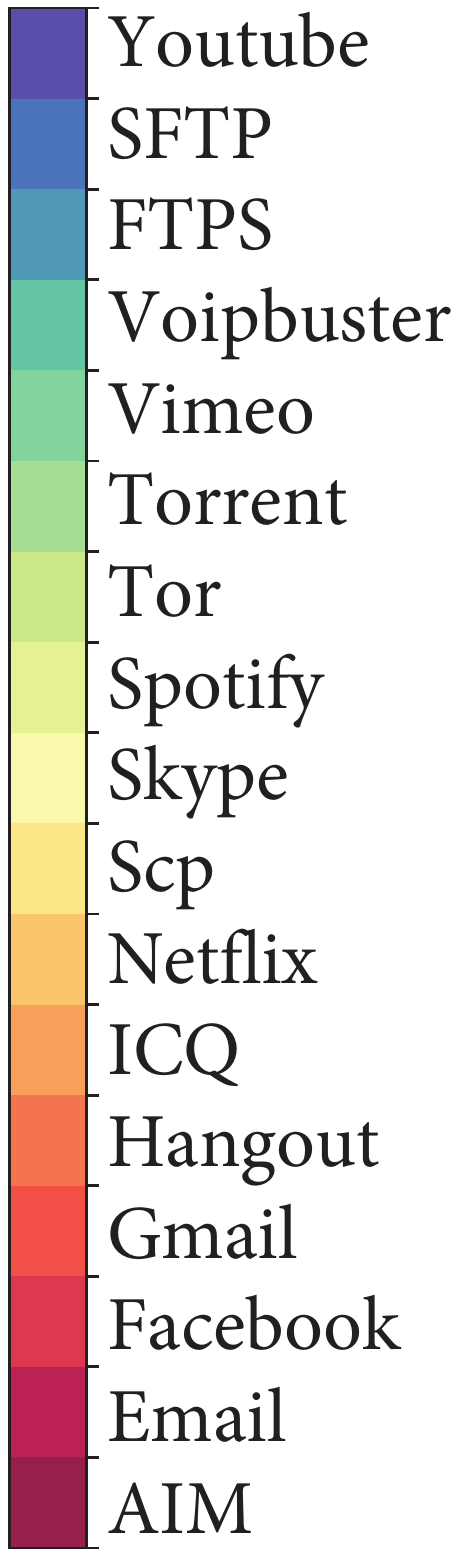}}
		
		\subfigure[Representation with PERT]{
			\includegraphics[width=0.24\linewidth,height = 0.2\textwidth]{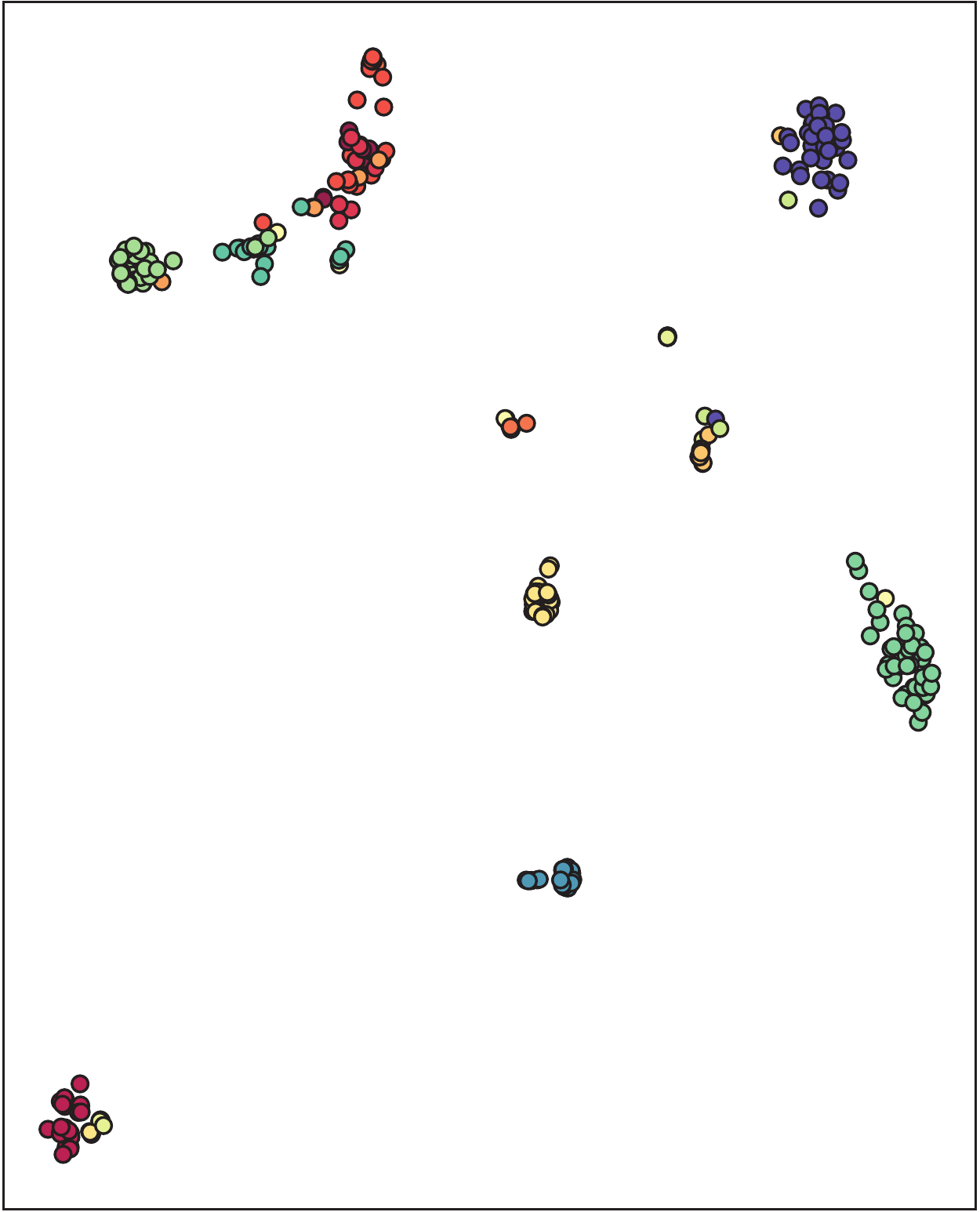}}
		\quad
		\subfigure[Representation with Transformer at flow]{
			\includegraphics[width=0.25\linewidth,height = 0.2\textwidth]{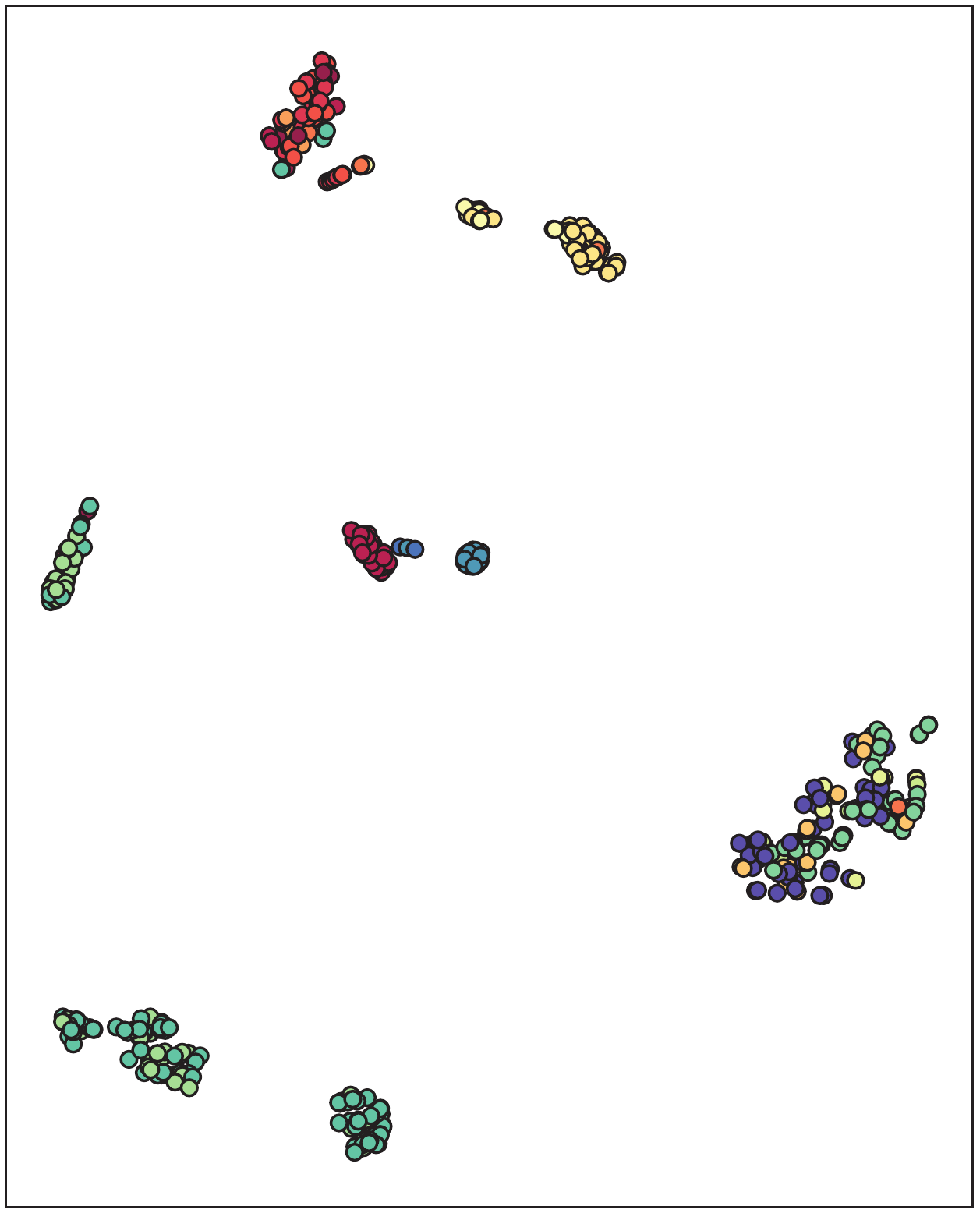}}
		\quad
		\subfigure[Representation with DF]{
			\includegraphics[width=0.24\linewidth,height = 0.2\textwidth]{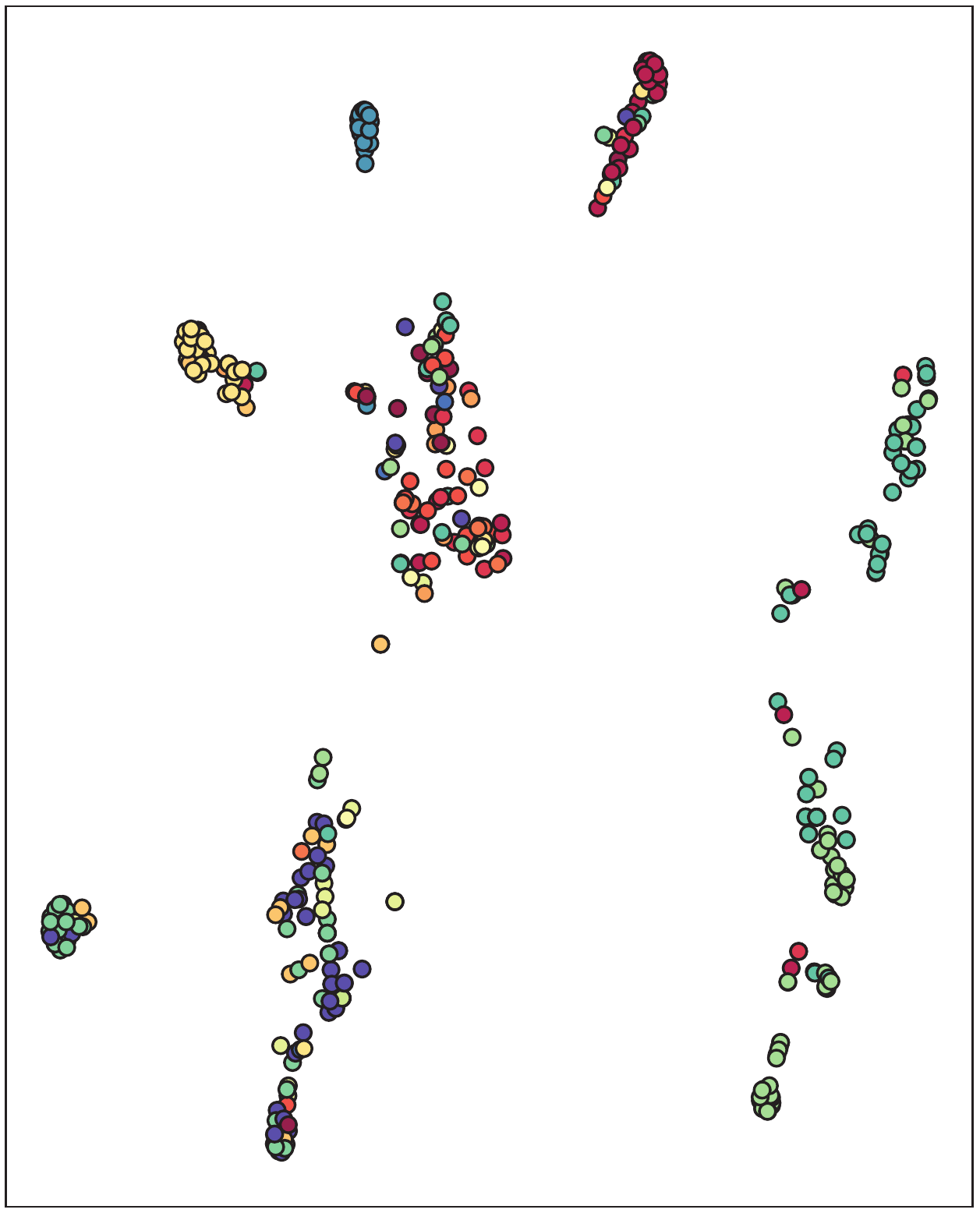}}
		\subfigure{
			\includegraphics[width=0.05\linewidth,height = 0.2\textwidth]{Figure-5-7.pdf}}
		\caption{t-SNE Visualization of Classification Boundaries with 6 Methods on ISCX-VPN-App Testset.}
		\label{fig-5}
	\end{figure*}
	\subsection{Qualitative Analysis}
	\label{qa}
	
	To further evaluate the performance differences between the models, we select 5 models for comparative analysis with ET-BERT, including Transformer at flow level, Transformer at packet level, DF, Deeppacket and PERT. The Transformer model as the baseline of ET-BERT can visually compare the improvement of our pre-trained model, and the remaining three models are representative methods to compare the prominence of our models.\par
	We use t-distributed stochastic neighbor embedding (t-SNE) to downscale the test set samples predicted by each model and plot them as two-dimensional images, as in Figure \ref{fig-5}. We choose the sampled ISCX-VPN-App dataset in Section \ref{ablation} and then show the best results for each model: (a-c) are packet-level results and (e-g) are flow-level results.\par
	There is no doubt that our model exhibits the best classification performance because ET-BERT captures patterns that can distinguish between different encrypted traffic even under the more secure new encryption protocols. Also with ET-BERT at the packet level, (b) and (c) fail to accurately classify applications especially AIM, ICQ and Gmail, which are used for online chat and Gmail provides online chat service in addition to email service. At the flow level, the classification effect of (f) and (g) is confusing, as YouTube and other streaming applications including Vimeo, Netflix and Spotify cannot be distinguished by these methods, while PERT performs relatively better but still suffers from the interference of applications with the same services.
	
\end{document}